\documentclass{article}
\usepackage{amsmath,amssymb,amsthm}
\ifx\pdfoutput\undefined
\usepackage[dvips]{graphicx}
\else
\usepackage[pdftex]{graphicx}
\usepackage[hyperindex]{hyperref}
\fi

\def\C{\mathbb{C}}
\def\R{\mathbb{R}}

\def\Z{\mathbb{Z}}
\def\Q{\mathbb{Q}}

\def\P{\mathbb{P}}
\def\Id{\hbox{1\kern-2.7pt I}}
\def\rank{{\rm rank}}

\newtheorem{thm}{Theorem}
\newtheorem{lem}[thm]{Lemma}
\newtheorem{cor}[thm]{Corollary}
\theoremstyle{definition}
\newtheorem{defn}[thm]{Definition}
\newtheorem*{rem}{Remark}
\newtheorem*{exampl}{Example}
\begin{document}
\author{Alexandre Stefanov}
\date{May, 2006}
\title{Finite orbits of the braid group action on sets of reflections}
\date{September, 2004}
\maketitle
\tableofcontents

\section{Introduction}

The problem of finding all finite orbits of the braid group action on tuples of
reflections appeared in the classification of semi-simple Frobenius manifolds.
These orbits correspond to algebraic solutions to equations of isomonodromic
deformations.

Suppose there is given a Fuchsian system of complex ordinary differential equations
\begin{equation}\label{Fuchs}
\frac{dy}{dz}=\sum_{i=1}^n\frac{A_i}{z-x_i}y,
\end{equation}
where $y(z)$ is a column vector of $m$ functions and $A_i$ are constant matrices.
Around each point $x\ne x_i$ there exist $m$ linearly independent solutions
$y_1,y_2,\dotsc,y_m$ and all solutions can be expressed by linear combinations of them.
It is known for linear systems that these solutions can be continued analytically
along any path, which doesn't pass through singularities of the coefficients.
For convenience such $m$ linear independent solutions are arranged into
a square matrix called the fundamental system of solutions.
For two fundamental systems $Y_1$ and $Y_2$2 the product $Y_1^{-1}Y_2$ is constant,
whence $Y_2=Y_1G$ for some $G\in GL(m,\C)$.
The result of the analytic continuation of a fundamental system of solutions
$Y(z)$ along a loop $\gamma$ based at $x$ will be another fundamental
system of solutions
\begin{equation}
Y_\gamma=YM_\gamma,
\end{equation}
where $M_\gamma$ is an invertible matrix depending only on the homotopy class
of $\gamma$. This gives us a linear representation of the fundamental group
\begin{equation}
\pi_1(\C\setminus\{x_1,\dotsc,x_n\})\to GL(m,\C)
\end{equation}
called the monodromy representation. Because of the freedom in the choice of the
fundamental system $Y_2=Y_1G$, $Y_{2(\gamma)}=Y_2M_{2(\gamma)}$,
$M_{2(\gamma)}=G^{-1}M_{1(\gamma)}G$, the monodromy representation is fixed
by the Fuchsian system only up to conjugation.

Deformations of the singularity points $x_i=x_i(t)$ and the matrix residues
$A_i=A_i(t)$ preserving the monodromy up to conjugation are called
isomonodromic deformations. These obey Schlesinger's equations
\begin{equation}
\frac{\partial A_i}{\partial x_j}=\frac{[A_i,A_j]}{x_i-x_j},\ i\ne j\quad
\frac{\partial A_i}{\partial x_i}=-\sum_{j\ne i}\frac{[A_i,A_j]}{x_i-x_j}.
\end{equation}
In a geometric language, as it is explained in~\cite{Ph,Var}, the above equations
define a non-linear flat connection on the fibre bundle
\begin{equation}
{\cal M}^*:=(O_1\times\dotsb O_n\times O_\infty)/GL_m(\C)\times B\to B
\end{equation}
over $B:=\C^n\setminus\{x_i=x_j,i\ne j\}$, where $O_i$ is the adjoint orbit
of $A_i$, which is preserved by the equations and $A_\infty=-\sum_{i=1}^n A_i$
is the residue at infinity. On the other hand the deformed
equations~\eqref{Fuchs} define the fibre bundle
\begin{equation}
M:={\rm Hom}(\pi_1(\C\setminus\{x_1,\dotsc,x_n\}),GL_m(\C))/GL_m(\C)\to B,
\end{equation}
equipped with a complete flat connection, defined locally by identifying
representations taking the same values on a fixed set of generators of the
fundamental group. The isomonodromy connection of Schlesinger's
equations is the pull-back of the natural bundle map ${\cal M}^*\to M$ from
the Fuchsian system to its monodromy representation. The monodromy of the
connection on $M$ and correspondingly on ${\cal M}^*$ amounts to an action
of the fundamental group of $B$ on the fibres, which is the braid group action
on the monodromy data
\begin{equation}
\sigma_i:(M_1,\dotsc,M_n)\mapsto(M_1,\dotsc,M_{i-1},M_iM_{i+1}M_i^{-1},
M_{i+2},\dotsc,M_n),
\end{equation}
where the $\sigma_i$ are the standard generators of the braid group
(see~\cite{Bir}).

The solutions to isomonodromic deformation equations of two dimensional
Fuchsian systems with four singularities on the Riemannian sphere $\P^1$
can be expressed through
solutions of the Painlev\`e VI equation. In~\cite{DM} were found all algebraic
solutions to one-parameter family of Painlev\`e VI equations, which correspond
to the finite orbits of the braid group action on triples of reflections.
It was shown that such orbits correspond to pairs of reciprocal regular
polyhedra or star-polyhedra (see~\cite{Cox}).

Algebraic functions have finite number of branches, therefore, in order to find all
algebraic solutions of the isomonodromic deformation
equations one must find all finite orbits of the braid group action on tuples
of linear transformations under the equivalence of simultaneous conjugation.
One class of transformations for which this action is particularly simple is
that of reflections, since a generic $n$-tuple of reflections can be specified by
a square matrix, called here the arrangement matrix. It is the Gram matrix
of the normed eigenvectors with eigenvalue $-1$, provided there is a
non-degenerate symmetric bilinear form invariant under all reflections.
It is known also as the Cartan matrix when these vectors are simple roots in
a root system. The action of
the braid group on the entries of these matrices, however, is non-linear.
It was conjectured by Dubrovin~\cite{D1} that all finite orbits of the braid
group on $n$-tuples of reflections come from finite Coxeter groups. The orbits on
non-redundant generating reflections in finite Weyl groups were found 
in~\cite{Voigt} and it was shown that these are in one-to-one correspondence
with the conjugacy classes of quasi-coxeter elements in these groups.

In the present article Dubrovin's conjecture is proved. Moreover, it is shown
that all finite orbits on singular matrices come from redundant generators
in finite Coxeter groups. Such matrices, however, represent non-unique
equivalence classes of $n$-tuples of reflections, and if the corank is greater than
one some of these equivalence classes depend on additional continuous parameters.
The question when the orbits on these parameters are finite is not considered
here.

The approach in this article is combinatorial. There are found universal sets
of generating reflections in each finite Coxeter group. These sets possess
maximal symmetry, and, fortunately, all conjugacy classes of quasicoxeter elements
are obtained from their products taken with the possible inequivalent orderings.
The universal sets allow inductive construction of all symmetric arrangement
matrices in finite orbits of the braid group using only the classification
of finite orbits on triples of reflections. In the course of this construction
we recover all finite Coxeter groups without using the standard generators
corresponding to reflections on the walls of Weyl chambers.

Another viewpoint for our construction are Schwarz triangles and their
higher dimensional analogues. The elements of the finite orbits on triples of
reflections with invertible Gram matrix are reflections, whose reflecting
planes intersect the sphere $S^2$ in Schwarz triangles. In the same way the
the reflecting hyperplanes of $n$-tuples of reflections in finite orbits
of the braid group action intersect the sphere $S^{n-1}$ in Schwarz simplexes,
which fit on a finite covering of the sphere by reflections on their sides.
The most symmetric Schwarz simplexes correspond to our universal sets of
generating reflections. For example the universal Schwarz simplex for the
group  $A_n$ is the projection of a face of the regular simplex on the
circumsphere and the angle between any two of its sides is $\frac{2\pi}{3}$
in contrast to the spherical simplex of the Weyl chamber of $A_n$, in which
the sides can be ordered so that the consecutive sides to meet at angle
$\frac{\pi}{3}$ and the non-consecutive to be orthogonal. This combinatorial
information is read directly from Coxeter-Dynkin diagram. We will
widely use diagrams to represent Schwarz simplexes and sets of reflections.

The classification of finite orbits of the braid group action on $n$-tuples
of reflections with invertible Gram matrix in the real Euclidean space was
done by Humphreys in~\cite{Hum1}. It coincides with ours, except for
the group $D_n$ for which the correct answer for the number of orbits is the
whole part of $n/2$ instead of $n-2$.
After the first preprint appearance of the present article, another very short
proof was found in~\cite{JM} and a flaw in the proof of Humphreys was pointed out.
Our assumptions are weaker than both these works as we consider not only
positive definite but arbitrary symmetrizable Gram matrices with complex entries.
In this way we treat simultaneously all linear groups generated by reflections.
This includes all Coxeter groups and also some non-Coxeter groups whenever
the corank of the Gram matrix is greater than one. The last can be
interpreted as groups of quasi-symmetries of almost periodic structures, or
as unfaithful representations of Coxeter groups satisfying additional
non-coxeter relations.

We require finiteness of the orbits of the braid group on only the equivalence
classes of ordered sets of reflections
\begin{equation}
(r_1,\dotsc,r_n)\sim (Gr_1G^{-1},\dotsc,Gr_nG^{-1})
\end{equation}
and not on the reflections themselves.

We show that all such orbits can be obtained from the (possibly redundant)
generators in finite Coxeter groups, provided the equivalence classes can be
specified by the arrangement matrix without additional continuous parameters.
This is the case for corank less than 2 arrangement matrices, and for
two extremal realizations of the matrices with greater corank namely
those in which the eigenvectors with eigenvalue $-1$ span a subspace of dimension
equal either to the rank of the arrangement matrix or to its size.

The actual classification of these orbits is done only for the invertible
Gram matrices. The characteristic polynomials of quasicoxeter elements are
calculated for each orbit, which is a new result for the icosahedral groups.
In another article~\cite{Al} we classify the orbits in the other extremal case
of rank 2 arrangement matrices. In this case the action can be linearized and
the obtained linear representation of the braid group coincides with the one
considered by Arnol'd in~\cite{Arn} for the odd number of reflections, while for even
number our representation is reducible and the nontrivial irreducible
component of it coincides with Arnol'd representation.

The result obtained here is not restricted only to sets of reflections. By
simple multiplication by $-1$ the reflections turn to half-turns to which
the same result applies. More subtle is the connection with tuples of
transvections, which are the nontrivial linear unipotent transformations
preserving point-wise hyperplanes of codimension one. An ordered set of such
transformations can be specified again by an arrangement matrix with zeroes on
the diagonal. If these transvections preserve a non-degenerate alternating
form, the arrangement matrix can be taken antisymmetric. The action of the
braid group on these antisymmetric matrices coincides with the action on
symmetric ones. This duality was used in~\cite{DM,D2,Ph} to switch from one
picture to the other using Laplace transformation to convert the
Fuchsian system with monodromy generated by transvections into a system
with one regular and one irregular singularity
\begin{equation}\label{irreg}
\frac{dY}{dz}=\left(U+\frac{V}{z}\right)Y\,,\quad z\in\C
\end{equation}
then applying a scalar shift and converting back
to a Fuchsian system with monodromy generated by reflections. The monodromy
data for the system~\eqref{irreg} is given by Stokes matrices, which relate
analytic solutions having the same asymptotic expansion in different sectors
centered at the irregular singular point $z=\infty$. In this case there is
essentially one Stokes matrix, and,  in an appropriate basis, it is upper
triangular with ones on the diagonal. For this system there is a notion of
isomonodromic deformation, in which the parameters of deformation are the
pairwise distinct eigenvalues of the matrix $U$. The fundamental group of
the space of parameters of the isomonodromic deformation is again the braid
group with $n$ strands, which gives an action of this group on the Stokes matrices.
The action of the braid group on Stokes matrices is the same
nonlinear action as for the symmetric and antisymmetric matrices. Yet another
place where the same action appears is the helix theory, where it is called
the braid group action on semi-orthonormal bases~\cite{Gor}. In the last case,
the entries of the upper triangular matrices are integer because it is
a cohomology theory.

\section{Action of the braid group}

Let's consider the free group ${\cal F}_n$ with $n$ generators as the fundamental
group $\pi_1(\C\setminus\{p_1,p_2,...,p_n\},O)$ of the (complex) plane with $n$
points removed, with some fixed base point $O$. The generators of ${\cal F}_n$ are
elementary cycles around points $p_k$.

The braid group ${\cal B}_n$ on $n$ strands can be defined as the group of homotopic
classes of diffeomorphisms of the plane with $n$ points removed
${\fam0\it Diff}(\C\setminus\{p_1,p_2,...,p_n\})$. It is generated by
$n-1$ standard generators $\sigma_1,\sigma_2,\dotsc,\sigma_{n-1}$ subject to
the following generating relations (see \cite{Bir})
\begin{equation}\label{braidrel}
\left\{\begin{array}{rl}
\sigma_i\sigma_{i+1}\sigma_i&=\sigma_{i+1}\sigma_i\sigma_{i+1}\\
\sigma_i\sigma_j&=\sigma_j\sigma_i,\quad j\ne i\pm1.\end{array}\right.
\end{equation}
In this context, it is evident that each braid transforms the cycles
$\gamma:S^1\to\C\setminus\{p_1,p_2,...,p_n\}$, and homotopically equivalent
diffeomorphisms transform a given cycle into homotopically equivalent cycles.
Therefore we have a natural inclusion ${\cal B}_n\subset {\mathcal Aut}({\cal F}_n)$
\begin{equation}
{\cal B}_n=\pi_0({\fam0\it Diff}(\C\setminus\{p_1,p_2,...,p_n\})):
{\cal F}_n\to {\cal F}_n =\pi_1(\C\setminus\{p_1,p_2,...,p_n\}) .
\end{equation}

\begin{figure}[htb]
\begin{center}\includegraphics{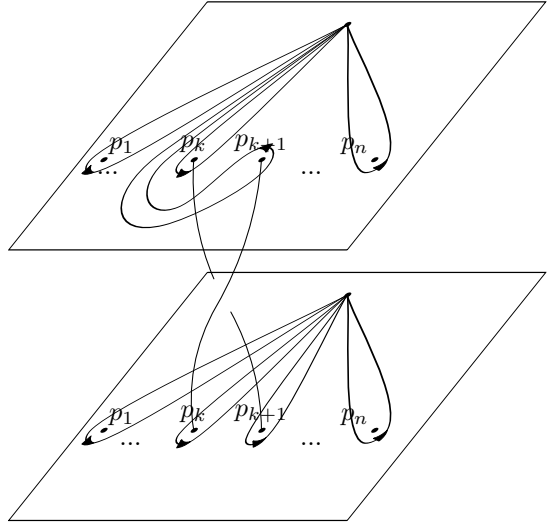}\end{center}
\caption{Hurwitz action on the fundamental cycles}\label{fig1}
\end{figure}
Under convention, the generators of ${\cal F}_n$ and ${\cal B}_n$ to be as
in~Fig.\ref{fig1}, we have
\begin{equation}
\label{braid}
\begin{array}{rcl}
\sigma_k(g_k) &=&g_kg_{k+1}g_k^{-1}\\
\sigma_k(g_{k+1}) &=&g_k\\
\sigma_k(g_l) &=&g_l\quad l\neq k,k+1\,.\\
\end{array}
\end{equation}

This action can be considered also over any ordered set of elements of a 
group. Our aim is to classify all the finite
orbits of ${\cal B}_n$ on groups generated by reflections. Throughout this
paper the braids act from the left
$(\sigma_1\sigma_2)(g1,\dotsc,g_n)=\sigma_1(\sigma_2(g1,\dotsc,g_n))$.

\subsection{Braid group action on arrangements of reflections}\label{reflsec}

\begin{defn}
Reflection in a linear space $V$ of arbitrary dimension
over the field of complex numbers $\C$, is a linear
transformation of period $2$ fixing point-wise hyperplane of codimension $1$.
\end{defn}
The general form of a reflection is
\begin{equation}
r=\Id - v\otimes v^\vee\quad\quad v\in V\setminus\{0\},
v^\vee\in V^*\setminus\{0\}\quad \langle v,v^\vee\rangle=v^\vee(v)=2,
\end{equation}
where the elements of the tensor product $V\otimes V^*$ are naturally identified
with endomorphisms of $V$, and $\Id$ is the identity operator.

The pair $(v,v^\vee)$ is unique for the reflection up to the change 
$(v,v^\vee)\mapsto (\lambda v,\lambda^{-1}v^\vee)$.
Given $n$ reflections
$r_i=\Id-v_i\otimes v_i^\vee\,,$ $v_i^\vee(v_i)=2$,
their relative position can be characterized by the arrangement matrix
\begin{equation}
B_{ij}= v_i^\vee(v_j)\,,
\qquad r_i=\Id-v_i\otimes v_i^\vee\,,\quad B_{ii}=v_i^\vee(v_i)=2\,.
\end{equation}

The same reflections can be characterized by different matrices $B$ and $B'$ if
\begin{equation}
\label{freed}
B_{ij}=\lambda_i{\lambda_j}^{-1}B'_{ij}
\end{equation}
for some nonzero numbers $\{\lambda_i\}_{i=1}^n$. In~\eqref{freed} appear only
ratios of the numbers $\lambda_i$ so one may always fix $\lambda_1=1$.
The equation~\eqref{freed} defines an equivalence relation on the arrangement matrices.
An arrangement matrix $B$ will be called {\it symmetrizable} if it is equivalent
to a symmetric matrix.
If the reflections preserve a non-degenerate quadratic form, the matrix
$B_{ij}$ is symmetrizable. We will consider only symmetrizable arrangement
matrices but will not assume the existence of invariant symmetric bilinear form.
The arrangement matrices of simple reflections in Coxeter groups are known as
Cartan matrices.

Let's remark that we do not set any restrictions on these reflections.
Usually it is required that the group of reflections should act properly
discontinuously on some geometric space. For vector spaces only the finite
groups of reflections act in this way. The affine and hyperbolic Coxeter
systems act properly discontinuously on affine and hyperbolic spaces
correspondingly. Allowing this greater freedom in the arrangement of
generating reflections permits us to include some non-Coxeter groups.
Moreover this gives us a uniform way to deal with Coxeter groups, because
the $n$-dimensional affine or hyperbolic spaces can be embedded in a 
$(n+1)$-dimensional vector space, where the reflections of the one are
reflections of the other.

The braid group transforms equivalence classes of arrangement
matrices $B$ as well as ordered sets of reflections.
So we have an action of
the braid group on an ordered $n$-tuple of reflections and on the equivalence
classes of arrangement matrices. Finite orbits of reflections imply
finite orbits of configuration matrices called onward $B$-orbits,
but the opposite isn't necessarily true. As it was stated before we will
consider only symmetric configurations $B_{ij}=B_{ji}$. The freedom
(\ref{freed}) for symmetric matrices is restricted:
\begin{equation}
\label{freed1}
B'_{ij}=\frac{\lambda_i}{\lambda_j}B_{ij}=B'_{ji}=
\frac{\lambda_j}{\lambda_i}B_{ij}\quad\Rightarrow\quad
\lambda_i^2=\lambda_j^2=\lambda_1^2=1\quad\Rightarrow\quad
\lambda_i=\pm 1
\end{equation}

The action of the braid group on the ordered sets of reflections induces
action on the space of symmetric arrangement matrices given by
\begin{equation}
\begin{array}{lcl}
{[\sigma_i(B)]}_{i\,j}&=&B_{i+1\,j}-B_{i\,i+1}B_{i\,j}\,,\quad j\ne i,i+1\\
{[\sigma_i(B)]}_{i+1\,j}&=&B_{ij}\,,\quad j\ne i\\
{[\sigma_i(B)]}_{i\,i+1}&=&-B_{i\,i+1}\\
{[\sigma_i(B)]}_{j\,j}&=&B_{j\,j}=2\\
{[\sigma_i(B)]}_{k\,j}&=&B_{k\,j}\,,\quad k\neq i,i+1\,.
\end{array}
\end{equation}
These transformations can be written in a compact form
\begin{equation}\label{sym}
\sigma(B)=K_\sigma(B)\cdot B\cdot K_\sigma(B)\,,
\end{equation}
where the symmetric matrices $K_\sigma(B)$ for the canonical generators of
the braid group are
\begin{equation}
(K_{\sigma_i}(B))_{j\,k}=\delta_{j\,k}-\delta_{i\,j}\delta_{j\,k}(1+B_{i\,i+1})
-\delta_{i+1\,j}\delta_{j\,k}+\delta_{i\,j}\delta_{i+1\,k}+
\delta_{i\,k}\delta_{i+1\,j}.
\end{equation}

The same action~\eqref{sym} can be defined on upper triangular matrices with ones
on the diagonal and it coincides with the action of the braid group on Stokes
matrices. Indeed the (anti-)symmetrization of Stokes matrices and the action of the
braid group commute
\begin{equation}
\sigma(S\pm S^T)=\sigma(S)\pm \sigma(S)^T.
\end{equation}

The antisymmetrized Stokes matrix $A=S-S^T$ can be interpreted as an
arrangement matrix of ``symplectic'' pseudo-reflections, preserving an
antisymmetric form of highest rank. A symplectic pseudo-reflection, called
also transvection, can be
defined as a linear transformation, fixing point-wise a hyperplane of
codimension 1 and having all eigenvalues equal to one. These requirements
imply that the Jordan canonical form of such transformation is
\begin{equation}
p=\begin{pmatrix}1&1&0&\cdots&0\\
0&1&0&\cdots&0\\
\vdots&\ddots&1&\ddots&\vdots\\
\vdots&&\ddots&\ddots&0\\
0&\cdots&&0&1
\end{pmatrix}
\end{equation}
and it can be written as
\begin{equation}
p=\Id + v\otimes v^\vee\quad\quad v\in V\setminus\{0\},
v^\vee\in V^*\setminus\{0\}\quad v^\vee(v)=0.
\end{equation}

The relative positions of $n$ such pseudo-reflections is given by their
arrangement matrix
\begin{equation}
A_{ij}= v_i^\vee(v_j)\,,
\qquad p_i=\Id+v_i\otimes v_i^\vee\,,\quad A_{ii}=v_i^\vee(v_i)=0\,.
\end{equation}
If there is a preserved antisymmetric bilinear form, the matrix $A$ can
be taken antisymmetric. Note that for odd-dimensional cases, the antisymmetric
form is always degenerate so there is a subspace invariant under all
pseudo-reflections. This allows a reduction by one of the dimension of a
space in which an odd number of symplectic pseudo-reflections stay in generic
position. The transformations in the reduced space will be no more
pseudo-reflections.

\section{Presentations of arrangements of reflections}

As we have seen $n$ reflections in a linear space $V$ are determined by
$n$ pairs $(v_i,v_i^\vee)$. To such ordered sets of vectors and covectors
we associate an arrangement matrix $B_{ij}=v_i^\vee(v_j)$. Here we will
reconstruct the reflections from the arrangement matrix. We call this procedure
a {\it realization} of the matrix, and we aim to examine
how many essentially different realizations as arrangements of reflections allows
a given matrix $B$.

\subsection{Construction of an arrangement from its matrix}

We introduce notions of reducibility and decomposability of arrangement matrices
in a similar fashion to the theory of group representations.
\begin{defn}
Arrangement matrix $B$ is called decomposable if there is a
permutation matrix $\Lambda$ such that
\begin{equation}
\Lambda B\Lambda^{-1}=\begin{pmatrix}
B_{k\times k}^{(1)}&0\\0&B_{n-k\times n-k}^{(2)}\end{pmatrix}
\end{equation}
Otherwise the matrix is called indecomposable.
\end{defn}
\begin{defn}
Arrangement matrix $B$ is called reducible if there is a
permutation matrix $\Lambda$ such that
\begin{equation}
\Lambda B\Lambda^{-1}=\begin{pmatrix}
B_{k\times k}^{(1)}&0\\B_{n-k,k}^{(3)}&B_{n-k\times n-k}^{(2)}\end{pmatrix}
\end{equation}
Otherwise the matrix is called irreducible.
\end{defn}
For symmetrizable matrices both notions coincide.
It is easy to see that an arrangement of reflections $r_1,\dotsc,r_n$
will have decomposable matrix if and only if there is a proper subset
$S\subset\{r_1,\dotsc,r_n\}$ of reflections, commuting with the remaining ones
\begin{equation}
r_i\in S,r_j\not\in S\Rightarrow r_ir_j=r_jr_i.
\end{equation}
It is clear that the action of the braid group~\eqref{braid} will preserve
this property. In the classification of finite $B$-orbits we may restrict
our attention only to indecomposable matrices because a decomposable matrix will
have finite orbit only if its building indecomposable blocks have finite
orbits.

Given $n$ reflections as above we consider the group $G$ they generate. As
a linear group its presentation will be indecomposable if the arrangement is,
provided that there is not a subspace, on which the group $G$ acts
trivially. We will fix the space $V$ of the representation
to be the minimal possible i.e. we will avoid as much as possible the existence
of a subspace on which $G$ acts trivially without changing the reflections.

\begin{rem}
Although for a symmetric arrangement matrix the properties
irreducible and indecomposable coincide it isn't necessarily true for the
linear group $G$, which the reflections generate. For example
\begin{equation}
r_1=\begin{pmatrix}-1&0\\0&1\end{pmatrix},\quad
r_2=\begin{pmatrix}-1&0\\-1&1\end{pmatrix}
\end{equation}
are reflections and the group they generate consists of matrices of the form
\begin{equation}
g=\begin{pmatrix}\pm1&0\\k&1\end{pmatrix},\quad k\in\Z,
\end{equation}
and it is isomorphic to the infinite dihedral group $I_2(\infty)$.
Therefore the presentation is indecomposable but reducible. The arrangement
matrix for $r_1,r_2$ is also indecomposable
\begin{equation}
B=\begin{pmatrix}2&2\\2&2\end{pmatrix}.
\end{equation}
\end{rem}

With this example in mind we define minimality for $V$ by:
\begin{defn}
The linear space $V$ is minimal for the set of reflections $r_1,\dotsc,r_n$,
$r_i=\Id-v_i\otimes v_i^\vee$ 
specified by $n$ pairs $(v_i,v_i^\vee)$, $v_i\in V,v_i^\vee\in V^*,v_i^\vee(v_i)=2$,
if the following holds true for all vectors $v\in V$
\begin{equation}
\forall i\ v_i^\vee(v)=0\Rightarrow v\in{\rm span}(v_1,\dotsc,v_n).
\end{equation}
In other words every vector in $V$ must either be moved by at least one of the
reflections or be a linear combination of the given $n$ vectors.
\end{defn}

To any given arrangement of reflections in the space $V$ there is a
naturally associated dual arrangement in the dual space $V^*$ obtained by
exchanging the places of $v_i$-s and $v_i^\vee$-s.

\begin{lem}
$V$ is minimal for $(v_i,v_j^\vee)_{i,j\in N}$ if and only if $V^*$ is minimal for
$(v_i^\vee,v_j)_{i,j\in N}$
\end{lem}
\begin{proof}
Assume $V$ is not minimal so $\{v_j\}_{j\in J}\cup\{w_k\}_{k\in K}$ is a basis
in $V$ and there is some $w_k$ such that
$v_i^\vee w_k\equiv 0$. Now let
$\{u_j^\vee\}_{j\in J}\cup\{w_k^\vee\}_{k\in K}$
be the the dual basis in $V^*$.
$$v_i^\vee w_k\equiv0\Rightarrow v_i^\vee\in{\rm span}(\{u_j^\vee\},
\{w_l^\vee\}_{l\in K\setminus k})
\Rightarrow w_k^\vee\not\in {\rm span}(v_j^\vee).$$
$w_k^\vee v_j\equiv0$ hence $V^*$ is not minimal for $(v_i^\vee,v_j)$.
\end{proof}

We will do all calculations for arrangements of reflections in their minimal
space to avoid unnecessary complications.

\begin{defn}
Two arrangements $\{v_i\in V\},\,\{v_i^\vee\in V^*\}_{i=1,\ldots,n}$ and
$\{v'_i\in V'\},\,\{{v'_i}^\vee\in V'^*\}_{i=1,\ldots,n}$ are isomorphic if
there is an isomorphism $i:V\rightarrow V'$ mapping $v_i$ to $v'_i$ while
the pull-back $i^*:V'^*\rightarrow V^*$ maps ${v'_i}^\vee$ to $v_i^\vee$.
\end{defn}

\begin{lem}
A non-degenerate matrix $B$ allows up to isomorphism exactly one realization
as an arrangement of reflections.
\end{lem}
\begin{proof}

As $B_{ij}=v_i^\vee(v_j)$ is invertible, the vectors $v_j$ are linearly
independent and so are $v_i^\vee$. Let $N=\{1,\ldots,n\}$, and $\{v_i^\vee\}_{i\in N}
\cup\{w_k^\vee\}_{k\in K}$ be a basis in $V^*$. Denote the dual basis in $V$
by $\{u_i\}_{i\in N}\cup\{t_k\}_{k\in K}$.
\begin{equation}
v_j=\sum_{i\in N} a_{ji}u_i+\sum_{k\in K} a'_{jk}t_k
\end{equation}
\begin{equation}
v_i^\vee(v_j)=\sum_{i_1\in N} a_{ji_1}v_i^\vee(u_{i_1})+
\sum_{k\in K} a'_{jk}v_i^\vee(t_k)=a_{ji}=B_{ij}
\end{equation}
\begin{equation}
\Rightarrow v_j=\sum_{i\in N} B_{ij}u_i+\sum_{k\in K} a'_{jk}t_k
\end{equation}
\begin{equation}
\sum_{j\in N} B^{-1}_{ji_1}v_j=\sum_{i,j\in N} B^{-1}_{ji_1}B_{ij}u_i+
\sum_{j\in N,k\in K} B^{-1}_{ji_1}a'_{jk}t_k
\end{equation}
\begin{equation}
\Rightarrow u_i=\sum_{j\in N} B^{-1}_{ji_1}v_j -\sum_{j\in N,k\in K} B^{-1}_{ji_1}a'_{jk}t_k 
\end{equation}
therefore $\{v_j\}_{j\in N}$, $\{t_k\}_{k\in K}$ is also basis in $V$. By definition $v_i^\vee(t_k)\equiv 0$
and from the minimality of $V$ follows $t_k\in{\rm span}(v_j)$ which is
impossible. We conclude $K=\emptyset$.
\end{proof}

For degenerate matrices there appear several possibilities for non-iso\-morphic
realizations as $v_j$ could be linearly independent and $v_i^\vee$
linearly dependent with rank equal to the rank of $B$, or the opposite, or $v_j$-s could
be linearly dependent with greater rank than $B$.

The formal treatment in the remaining part of this section will be without assuming
symmetrizability of the arrangement matrices.
Let's denote by
$$B_{i\cdot}=(B_{i1},B_{i2},\ldots ,B_{in})\quad
B_{\cdot j}=(B_{1j},B_{2j},\ldots ,B_{nj})$$
the rows and columns of $B$. Let $\{B_{i\cdot}\}_{i\in I}$ be a basis in
${\rm span}(B_{i\cdot})$ for some $I\subset\{1,2,\ldots ,n\}$. This
subset $I$ is non-unique for $0<\rank(B)<n$.
In the same way $\{B_{\cdot j}\}_{j\in J}$ is a basis in the span of columns
of $B$ for some $J\subset N:=\{1,2,\ldots ,n\}$.
$|I|=|J|=r:=\rank (B)$, $\{v_i^\vee\}_{i\in I}$ and $\{v_j\}_{j\in J}$
are linearly independent.

\begin{equation}
B_{i\cdot}=\sum_{i_1\in I}a_{ii_1}B_{i_1\cdot}\quad{\rm if}\, i\in N\setminus I\qquad
B_{\cdot j}=\sum_{j_1\in J}b_{j_1j}B_{\cdot j_1}\quad{\rm if}\, j\in N\setminus J
\end{equation}

\begin{thm}
Any degenerate matrix $B$ of rank $r$ allows non-unique realization as an
arrangement of reflections. To specify a unique (up to isomorphism) arrangement
one must say which of the vectors $\{v_j\}_{j\in J''}$ and the covectors
$\{v_i^\vee\}_{i\in I''}$ are linearly independent. These sets $I''$ and $J''$
must include sets $I$ and $J$ of indices of rows and columns of $B$ forming
bases in the span of all rows and columns.
Such subsets $I'',J''$ may be chosen in $2^{2(n-r)}$ different ways.

Additionally one must fix $(n-|I''|)(|I''|-r)+(n-|J''|)(|J''|-r)$ arbitrary
constants in order to specify a unique realization of $B$. The dimension of
the minimal space for this arrangement is $|I''|+|J''|-r$.
\end{thm}

We have already chosen the sets $I,J$. As $\{v_i^\vee\}_{i\in I}$ are
linearly independent we may complement them by $\{v_i^\vee\}_{i\in I'}$
to a basis in ${\rm span}(v_i^\vee)$. There are $2^{n-r}$ possibilities for
the set $I'$. Analogously let $\{v_j\}_{j\in J\cup J'}$ form a basis
in ${\rm span}(v_j)$.

The remaining vectors and covectors are expressed through these
\begin{equation}
v_i^\vee=\sum_{i_1\in I\cup I'}a'_{ii_1}v_{i_1}^\vee\,,\quad i\in N\setminus(I\cup I')
\qquad
v_j=\sum_{j_1\in J\cup J'}b'_{j_1j}v_{j_1}\,,\quad j\in N\setminus(J\cup J')\,.
\end{equation}

The coefficients $a'_{ii_1}$ must satisfy
\begin{equation}\label{restr}
B_{ij}=v_i^\vee(v_j)=\sum_{i_1\in I\cup I'}a'_{ii_1}v_{i_1}^\vee(v_j)=
\sum_{i_1\in I\cup I'}a'_{ii_1}B_{i_1j}=
\sum_{i_1\in I}a_{ii_1}B_{i_1j}
\end{equation}
for $i\in N\setminus(I\cup I')\,\,j\in J$.

The matrix $\tilde B=(B_{ij})_{i\in I,j\in J}$ is invertible and for ease
of notation we will write $B^{-1}_{ji}$ instead of $\tilde B^{-1}_{ji}$.
Care must be taken as $B^{-1}_{ji}$ is defined only for $i\in I,j\in J$ and
\begin{align}
\sum_{j\in J}B_{i_1j}B^{-1}_{ji_2}&=
\begin{cases}\displaystyle\delta_{i_1i_2}& i_1\in I\\[5pt]
\displaystyle\sum_{j\in J,i_3\in I}a_{ii_3}B_{i_3j}
B^{-1}_{ji_2}=a_{i_1i_2}& i_1\in N\setminus I
\end{cases}\\
\sum_{i\in I}B^{-1}_{j_1i}B_{ij_2}&=
\begin{cases}\displaystyle\delta_{j_1j_2}& j_2\in J\\[5pt]
\displaystyle\sum_{i\in I,j_3\in J}B^{-1}_{j_1i}B_{ij_3}b_{j_3j_2}=
b_{j_1j_2}& j_2\in N\setminus J
\end{cases}.
\end{align}

After multiplying \eqref{restr} by $B^{-1}_{ji_2}$ and summing over $j\in J$
\begin{equation}
a_{ii_2}=\sum_{\substack{i_1\in I\cup I'\\j\in J}}{a'}_{ii_1}B_{i_1j}
B^{-1}_{ji_2}={a'}_{ii_2}+\sum_{i_1\in I'}{a'}_{ii_1}a_{i_1i_2}
\end{equation}
so
\begin{equation}
{a'}_{ii_2}=a_{ii_2}-\sum_{i_1\in I'}{a'}_{ii_1}a_{i_1i_2}\,.
\end{equation}
The coefficients ${a'}_{ii_1}\,i_1\in I'$ are independent and the remaining
${a'}_{ii_2}\,i_2\in I$ are calculated from them and the matrix $B$.

Let $\{v_i^\vee\}_{i\in I\cup I'}\cup\{w_k^\vee\}_{k\in K}$ be a basis in $V^*$.
Denote $\{u_i\}_{i\in I\cup I'}\cup\{w_k\}_{k\in K}$ the dual basis in $V$.
\begin{gather}
v_j=\sum_{i_1\in I\cup I'}c'_{i_1j}u_{i_1}+\sum_{k\in K}c_{kj}w_k\quad
v_i^\vee(v_j)=c'_{ij}=B_{ij}\\
\label{eq1}\Rightarrow
v_j=\sum_{i_1\in I\cup I'}B_{i_1j}u_{i_1}+\sum_{k\in K}c_{kj}w_k
\end{gather}
Multiplying by $B^{-1}_{ji_2}$ and summing over $j\in J$
\begin{equation}\label{eq2}
u_{i_2}=\sum_{j\in J}B^{-1}_{ji_2}v_j-\sum_{i_1\in I'}a_{i_1i_2}u_{i_1}-
\sum_{k\in K,j\in J}c_{kj}B^{-1}_{ji_2}w_k\,
\end{equation}
therefore $\{v_j\}_{j\in J},\{u_i\}_{i\in I'},\{w_k\}_{k\in K}$ is also
a basis in $V$.

After substituting \eqref{eq2} in \eqref{eq1} for $j\in J'$we obtain
\begin{equation}\label{wksp}
v_j=\sum_{j_1\in J}b_{j_1j}v_{j_1}+\sum_{k\in K}d_{kj}w_k
\end{equation}
where the coefficients $d_{kj}$ are obtained from $c_{kj}$.
The vectors $w_k$ were not fixed up to now so we may use any other basis in
${\rm span}(w_k)$. As $v_j\,,j\in J\cup J'$ are linearly independent by
assumption, $\tilde{w}_j=v_j-\sum_{j_1\in J}b_{j_1j}v_{j_1}$,
$j\in J'$ are linearly independent and belong to ${\rm span}(w_k)$ by~\eqref{wksp}.
We choose another basis in ${\rm span}(w_k)$ so that $w_j=\tilde{w}_j$, $j\in J'$
identifying a subset of $K$ with $J'$. Remark that nothing was said about the set $K$ till now, and we may assume it is a superset of $J'$.
Complete the basis in ${\rm span}(w_k)$ with $\{w_k\}_{k\in K\setminus J'}$. By assumption $v_i^\vee(w_k)\equiv0$ but 
$w_k\not\in{\rm span}(v_j)$ for $k\in K\setminus J'$ contradicting the
minimality condition on $V$ hence $K\setminus J'=\emptyset$.

\begin{cor}\label{realiz}
Given the matrix $B$ of size $n$ and rank $r$, the sets $I'',J''$ and the
constants $a'_{ii_1},b'_{j_1j}$ for $i_1\in I'',i\in N\setminus{I''}$,
$j_1\in J'',j\in N\setminus{J''}$, where $I''=I\cup I', J''=J\cup J'$; the
reflection arrangement can be build in the following way:

Let the basis vectors in the space $V^*$ be
$\{v_i^\vee\}_{i\in I''}$ and $\{w_k^\vee\}_{k\in J'}$. Denote the vectors of the
dual basis in $V$ by $\{u_i\}_{i\in I''},\{w_k\}_{k\in J'}$. The vectors
in the arrangement are
\begin{align}
v_j&=\sum_{i\in I}B_{ij}u_i,&j\in J\\
v_j&=\sum_{j_1\in J}b_{j_1j}v_{j_1}+w_j,&j\in J'\\
v_j&=\sum_{j_1\in J''}b'_{j_1j}v_{j_1},&j\in N\setminus{J''}\\
v_i^\vee&=\sum_{i_1\in I''}a'_{ii_1}v_{i_1}^\vee,&i\in N\setminus{I''}
\end{align}
\end{cor}

\begin{exampl}
The simplest example allowing demonstration of the above construction with most
of the features is the $3\times 3$ configuration matrix of rank~1:
\begin{equation}
B=\begin{pmatrix}2&2&2\\2&2&2\\2&2&2\end{pmatrix}
\end{equation}
We fix $I=J=\{1\}$. There are 16 variants to choose subsets
$I',J'\subset\{2,3\}$. If we let $I'=J'=\{2\}$ there will be 2 constants to
determine completely an arrangement. Calling them $a,b$ we have
\begin{equation}
v_3^\vee=(1-a)v_1^\vee+av_2^\vee,\qquad v_3=(1-b)v_1+bv_2.
\end{equation}
The dimension of the minimal space is $|I''|+|J''|-r=3$. We may take the standard basis vectors
in $\R^3$ to be $v_1,v_2,w$
\begin{equation}
v_1=(1,0,0)^T,\quad v_2=(0,1,0)^T,\quad w=(0,0,1)^T
\end{equation}
The covectors $v_1^\vee,v_2^\vee$ must obey the arrangement matrix. We take
\begin{equation}
v_1^\vee=(2,2,0),\quad v_2^\vee=(2,2,1)
\end{equation}
The three reflections in this arrangement are
\begin{gather}
r_1=\begin{pmatrix}-1&-2&0\\0&1&0\\0&0&1\end{pmatrix},\quad
r_2=\begin{pmatrix}1&0&0\\-2&-1&-1\\0&0&1\end{pmatrix},\\
r_3=\begin{pmatrix}2b-1&2b-2&ab-a\\-2b&1-2b&-ab\\0&0&1\end{pmatrix}
\end{gather}
This gives also an example of non-Coxeter group. One may easily check
that $r_ir_j$ has infinite period if $i\ne j$, but whenever $a=b$, the product
$r_1r_2r_3$ is a reflection therefore having period 2. Further analysis shows that all elements in this group
are either reflections or have infinite period. As an abstract group it has
a presentation
\begin{equation}
r_1^2=r_2^2=r_3^2=(r_1r_2r_3)^2=1
\end{equation}
which is clearly not Coxeter group (see~\cite{Bo,Hum}). It is isomorphic to
the group of congruence transformations on $\R$, preserving a set of points
with coordinates $x+y\sqrt{2}$, $x,y\in\Z$. This is easy to see as the
congruence
transformations on a line are only reflections and translations; and
three reflections about points with coordinates $0,1,\sqrt{2}$ generate
all such transformations.
\end{exampl}
\begin{rem}
The dimension of the minimal space of the reflection arrangements having a
given arrangement matrix $B$ of size $n$ and rank $r$ can be any integer from
$r$ to $2n-r$. We call the {\it minimal realization} this arrangement, whose minimal
space has dimension $r$. Analogously we call the {\it maximal realization} of the matrix
$B$ this arrangement of reflections, whose minimal space has dimension $2n-r$.
The minimal and maximal realizations are unique for every matrix $B$.
\end{rem}

In the applications we are interested, there is a non-degenerate bilinear form,
preserved by all reflections. As we have said in this case the matrix $B$
is symmetric

\begin{lem} Any the reflection $r$ preserving a non-degenerate 
symmetric bilinear form
$(\ ,\ )$ has the form $r=\Id-(v,\ )$, where $v$ is a vector
satisfying $(v,v)=2$.
\end{lem}
\begin{proof}
Writing as before $r=\Id -v\otimes v^\vee$ the invariance means
\begin{equation}
( u,w )= ( r(u),r(w) )=
( u,w )-( v^\vee(u)v,w )-( u,v^\vee(w)v )
+( v^\vee(u)v,v^\vee(w)v )
\end{equation}
for any pair of vectors $u,w$. If $v^\vee(u)=0$ and $v^\vee(w)\ne0$,
it must hold $( u,v^\vee(w)v)=0=( u,v )$.
Let us now substitute $w=u$ for a vector $u$, for which $v^\vee(u)\ne0$.
In this case $(v^\vee(u))^2( v,v )=2v^\vee(u)( v,u )$
therefore $v^\vee(u)=\frac{2( v,u )}{( v,v )}$
provided $( v,v )\ne0$. But if $( v,v )=0$,
it must hold $( v,u )=0$ for any $u$ which contradicts the
non-degeneracy of the form $(\ ,\ )$. Ii follows that always
$v^\vee=\frac{2( v,\ )}{( v,v )}$ and we may rescale $v$ to
make the denominator equal to 2.
\end{proof}

In this case the arrangement matrix $B_{ij}=v_i^\vee(v_j)=
( v_i,v_j)$ is the Gram matrix of the vectors $v_i$. To recover
the arrangement from this matrix we may use the above construction, taking
into account that now we have only $n$ vectors and a natural isomorphism
between $V$ and $V^*$.
\begin{equation}
v_i=\sum_{j\in I\cup I'}a'_{ij}v_j\quad B_{ik}=\sum_{j\in I}B_{jk}
\end{equation}
The following identities must hold
\begin{equation}
a'_{ij}=a_{ij}-\sum_{k\in I'}a'_{ik}a_{kj},\quad i\not\in I\cup I',j\in I
\end{equation}
so there are $(n-|I|-|I'|)|I'|$ independent parameters $a'_{ij}$ which must
be specified along with the matrix $B$ and the subsets $I,I'$ to fix a
unique set of reflections up to simultaneous conjugation.
There must be some additional vectors $\{w_j\}_{j\in I'}$, which together
with $\{v_i\}_{i\in I\cup I'}$ form a basis in the minimal space $V$.
A convenient choice of $w_j$ is one, for which
$(v_i,w_j)=\delta_{ij}$.

\subsection{Representation of the arrangement matrix by a graph}

At this point arises the question how transforms the arrangement matrix under
the action of the braid group in different realizations.
\begin{align}
r_ir_{i+1}r_i=(\Id-v_i\otimes v_i^\vee)(\Id-v_{i+1}\otimes v_{i+1}^\vee)(\Id-v_i\otimes v_i^\vee)
=\notag\\
\Id-(v_{i+1}-B_{ii+1}v_i)\otimes(v_{i+1}^\vee-B_{ii+1}v_i^\vee)
\end{align}
hence
\begin{equation}\label{matact}
\begin{array}{rcl}
\sigma_i(B)_{ij}&=&B_{i+1 j}-B_{i i+1}B_{ij}\\
\sigma_i(B)_{i+1,j}&=&B_{ij}\\
\sigma_i(B)_{i,i+1}&=&-B_{i,i+1}\\
\sigma_i(B)_{kj}&=&B_{kj}\,{\rm for}\, k\neq i,i+1\,.
\end{array}
\end{equation}

We see that although if $B$ is degenerate it can have different
non-iso\-morph\-ic realizations as reflection arrangements, it transforms
uniformly
by the braid group. This key observation will allow us later to classify
the finite orbits arising from finite groups as well as those arising from
infinite groups.

For the sake of visualization we will represent the matrix $B$ by a graph
$\Gamma$ with vertices $\nu_i$, $i=1,\ldots,n$ and labeled edges
$(\nu_i,\nu_j)\in{\rm Edge}(\Gamma)\Leftrightarrow B_{ij}\neq0$ with labels
$g(\nu_i,\nu_j)=\pm\frac{n}{k}\,$ if
$B_{ij}=\pm 2\cos\frac{\pi k}{n},\, 0<\frac{k}{n}<\frac{1}{2}$.
This restriction on the possible values in $B_{ij}$ is necessary when we
consider matrices from finite orbits of the braid group as we will see later.
Indecomposable arrangement matrices have connected graphs. In analogy with
Dynkin diagrams we omit the labels $g(\nu_i,\nu_j)=\pm 3$
and write only the sign. In contrast to Dynkin diagrams we will omit the
positive signs and write only the negative ones. Remember that the angles
between simple roots are non-acute therefore all non-diagonal entries
in a Cartan matrix are non-positive. We should always take into
account the identification~(\ref{freed}) of graphs and matrices representing
the same reflection configuration. In particular it makes redundant the signs
when the graph is a tree or more than one negative signs when the graph
contains one cycle. In graphs we always abbreviate $\frac{5}{2}$ to $5'$
as it is the only fraction to appear.

When investigating reflection arrangements generating given Coxeter group
there are considered certain ``universal'' graphs without indexing of the
vertices which will be called unindexed:
\begin{multline}
\Gamma=\{V,E,g\}, V=\{v_1,\cdots,v_n\},
E\subseteq\{\{v_i,v_j\},\,v_i,v_j\in V\},\\
g:E\to\{\pm\frac{n}{k}\}_{0<2k<n}.
\end{multline}

Indexing of the vertices is equivalent to their linear ordering:
\begin{equation}
\Gamma=\{V,E,g,\prec\}
\end{equation}
Two graphs, which differ only on ordering of their vertices will be
called {\it similar}.
By a subgraph $\Gamma'$ of the graph $\Gamma$ will be understood
\begin{multline}
\Gamma'=\{V',E',g',\prec'\}, V'\subseteq V,
E'=\{\{v_i,v_j\}\in E\,,v_i,v_j\in V'\},\\
g'=\left.g\right|_{E'},\prec'=\left.\prec\right|_{V'}. 
\end{multline}
We call the graph $\Gamma$ an extension of $\Gamma'$ by $\#V-\#V'$ vertices
and $\#E-\#E'$ edges.

Graphs corresponding to invertible arrangement matrices will be called
{\it non-degenerate} and those corresponding to singular arrangement matrices
-- {\it degenerate}.

\subsection{Invariants of the action of the braid group}

As it is seen in~(\ref{braid}) the element
\begin{equation}
C=g_1g_2\dotsm g_kg_{k+1}\dotsm g_n=
g_1g_2\dotsm g_kg_{k+1}g_k^{-1}g_k\dotsm g_n
\end{equation}
is an invariant of the action. For the canonical generators of Coxeter
groups $C$ is called Coxeter element and in our case of arbitrary set
of reflections it will be called quasicoxeter element.
As the matrix $B$ specifies only the relative positions of the reflections
$r_1,r_2\dotsc r_n$ we see that the conjugation class of $C$ discriminates
the different orbits of ${\cal B}_n$ acting on the matrix $B$.

We proceed with expressing $C$ by $B$
\begin{multline}
C=\prod_{i=1}^n(\Id-v_i\otimes v_i^\vee)=\Id-\sum_{k=1}^n\,\sum_{\scriptscriptstyle
1\le i_1<i_2<\dotsb
i_k\le n}(-1)^{k+1}v_{i_1}\otimes v_{i_1}^\vee(v_{i_2})v_{i_2}^\vee(v_{i_3})\dotsm
v_{i_{k-1}}^\vee(v_{i_k})v_{i_k}^\vee\\
=\Id-\sum_{k=1}^n\,(-1)^{k+1}\sum_{\scriptscriptstyle 1\le i_1<i_2<\dotsb
i_k\le n}B_{i_1i_2}B_{i_2i_3}\dotsm B_{i_{k-1}i_k}v_{i_1}\otimes v_{i_k}^\vee\,.
\end{multline}

The last expression may be simplified by the following trick
\begin{equation}
\sum_{\scriptscriptstyle i<i_1<i_2<\dotsb<i_k<j}B_{ii_1}B_{i_1i_2}\dotsm
B_{i_kj}=(U^{k+1})_{ij}\quad\text{where}\quad U_{ij}=
\begin{cases}B_{ij}&i<j\\
0&i\ge j\,.
\end{cases}
\end{equation}

The matrix $U$ is nilpotent so
\begin{equation}\label{quasicox}
C=\Id-\sum_{i,j=1}^n\sum_{k=0}^\infty(-1)^k{U^k}_{ji}v_j\otimes v_i^\vee=
\Id-\sum_{i,j=1}^n(\delta+U)_{ji}^{-1}v_j\otimes v_i^\vee\,.
\end{equation}

If the matrix $B$ is non-degenerate $\{v_j\}$ form a basis in $V$. Denoting the
dual basis in $V^*$ by $\{u_j^\vee\}$ we have
\begin{equation}
C=\sum_{i,j=1}^n\left[\delta-(\delta+U)^{-1} B\right]_{ji}v_j\otimes u_i^\vee
\end{equation}
Introducing $V=B-U$ and expressing $B$ through $U,V$ we obtain
\begin{equation}\label{quasicox1}
C=\sum_{i,j=1}^n\left[(\delta+U)^{-1}(\delta-V)\right]_{ji}v_j\otimes u_i^\vee\,.
\end{equation}

Whenever rank$(B)=r<n$ the basis in $V^*$ is $\{v_i^\vee\}_{i\in I\cup I'}\cup
\{w_j^\vee\}_{j\in J'}$ and the dual basis in $V$ is $\{u_i\}\cup\{w_j\}$.
\begin{equation}
v_i^\vee=\sum_{i_1\in I\cup I'}a'_{ii_1}v_{i_1}^\vee\quad v_j=
\begin{cases}
\sum_{i_1\in I\cup I'}B_{i_1j}u_{i_1}&j\in J\\
w_j+\sum_{i_1\in I\cup I'}B_{i_1j}u_{i_1}&j\in J'\\
\sum_{\substack{i_1\in I\cup I'\\j_1\in J\cup J'}}B_{i_1j_1}b'_{j_1j}u_{i_1}
+\sum_{j_1\in J'}b'_{j_1j}w_{j_1}&\text{else.}
\end{cases}
\end{equation}
To write a compact formula it is best to extend the definition of $a'_{ii_1}$,
$b'_{j_1j}$ to all subscripts by
\begin{equation}
a'_{ii_1}=\begin{cases}\delta_{ii_1}&i\in I\cup I'\\0&i_1\not\in I\cup I'
\end{cases}\qquad b'_{j_1j}=
\begin{cases}\delta_{j_1j}&j\in J\cup J'\\0&j_1\not\in J\cup J'\,.
\end{cases}
\end{equation}

Substituting $v_i^\vee$, $v_j$ in \eqref{quasicox} we finally obtain
\begin{multline}
C=\sum_{i_1i_2\in I\cup I'}(\delta-Bb'(\delta+U)^{-1}a')_{i_1i_2}
u_{i_1}\otimes v_{i_2}^\vee+\sum_{j\in J'}w_j\otimes w_j^\vee\\
-\sum_{\substack{j\in J'\\i\in I\cup I'}}\left(b'(\delta+U)^{-1}a'\right)_{ji}
w_j\otimes v_i^\vee\,.
\end{multline}

\section{Two and three reflections}

From this point on we will consider only symmetric arrangement matrices.
The first non-trivial case to study are the orbits of the braid group
action on configuration matrices of three reflections as the action is trivial
on $2\times 2$ matrices.
\begin{equation}
 B=\begin{pmatrix}2&a&b\\a&2&c\\b&c&2\end{pmatrix}
\end{equation}

The canonical generator $\sigma_1$ of the braid group act on this matrix by
\begin{equation}
\sigma_1(B)=\begin{pmatrix}2&a&ab-c\\a&2&b\\ab-c&b&2\end{pmatrix}\,.
\end{equation}
This is a linear transformation on the pair $b,c$ and must have finite period
if the orbit of ${\cal B}_3$ is finite. This will take place only if the eigenvalues
of $\bigl(\begin{smallmatrix}a&-1\\1&0\end{smallmatrix}\bigr)$ are roots of
unity so expressing $a$ we obtain
\begin{equation}
a=2\cos\alpha\,,\quad \alpha\in\pi\Q\,.
\end{equation}

As it can be seen in \eqref{braid}, the braid $\sigma_1\sigma_2\dotsm
\sigma_{n-1}$ permutes cyclically $g_1,g_2,\dotsc, g_n$ and conjugates them
by $g_1$. The configuration matrix $B$ remains unchanged upon simultaneous
conjugation of the reflections therefore the braid $\sigma_1\sigma_2$ permutes
$a,b,c$ cyclically. It follows that all $a,b$, and $c$ must be twice the cosines of
rational parts of the straight angle.

If the matrix $B$ is degenerate it must have the form
\begin{equation}
B=\begin{pmatrix}2&2\cos\alpha&2\cos\beta\\
\cos\alpha&2&2\cos(\alpha+\epsilon\beta)\\
2\cos\beta&2\cos(\alpha+\epsilon\beta)&2\end{pmatrix},\quad\epsilon=\pm1.
\end{equation}
Changing $\beta\mapsto -\beta$ we may assure $\epsilon=-1$. The generators
of the braid group transform parameters $\alpha,\beta$ by
\begin{equation}
\sigma_1:\left\{\begin{array}{rl}\alpha&\mapsto\alpha\\
\beta&\mapsto\alpha+\beta\end{array}\right.\quad
\sigma_2:\left\{\begin{array}{rl}\alpha&\mapsto\alpha-\beta\\
\beta&\mapsto\alpha\end{array}\right..
\end{equation}

Returning to the initial parameters $a,b,c$ we must identify $\alpha,\beta$
modulo $2\pi$. One may see that the condition $\alpha,\beta\in\pi\Q$ is
sufficient for the orbit of ${\cal B}_3$ to be finite. Such matrix represents,
in the minimal realization, a redundant set of generators of some finite
dihedral group $I_2(k)$.

The case of non-degenerate symmetric $3\times 3$ matrix $B$ is considered in~\cite{DM}
where it was proved that the only finite orbits come from matrices representing
configurations of reflections generating finite three-dimensional
Coxeter groups. There is a one-to-one correspondence between the conjugacy
classes of quasicoxeter elements in them and orbits of the braid group.
Even more appealing is the correspondence between the orbits and the pairs
of reciprocal regular polyhedra and star-polyhedra in the three dimensional
space (\cite{Cox1}).

Summarizing, the finite orbits of the braid group on $3\times 3$ configuration
matrices represent reflections, generating finite groups. In case of degenerate
matrix there is a realization of it where one of the reflections belongs to the
group generated by the other two.

In the classification of all finite orbits of the braid group action on
arrangement matrices we will follow an inductive procedure for which the next
lemma is essential. We always identify a graph with the arrangement matrix it
represents.

\begin{lem}\label{subgraph}
A graph $\Gamma$ containing a subgraph $\Gamma'$ which has infinite orbit under
the action of the braid group has an infinite orbit itself.
\end{lem}
\begin{proof}
Let the vertices of $\Gamma$ be numbered $1,2,\dotsc,n$ and those of $\Gamma'$
when ordered $1\le i(1)<i(2)<\dotsb<i(k)\le n$. The braid $\sigma_{n-1}
\sigma_{n-2}\dotsm\sigma_{i(k)}$ moves the $i(k)$-th reflection to the last
position. Acting by $\sigma_{n-2}\sigma_{n-3}\dotsm\sigma_{i(k-1)}$ will
bring the $i(k-1)$-th reflection to next to the last position leaving
the last unchanged. Continuing in the same manner we may bring all vertices
of $\Gamma'$ to consecutive numbers without changing the subgraph $\Gamma'$.
It is clear that a subgroup of ${\cal B}_n$ will have an infinite orbit when acting
on the obtained graph.
\end{proof}

In the remaining part we will investigate of the finite orbits along the following lines:
\begin{itemize}
\item First are found the orbits of the braid group on non-redundant
sets of generators in finite Coxeter groups. These are necessarily finite
because there are only finite number of combinations of generators in a finite
group. As it is not obvious how to find all possible such sets we use an
inductive argument: choosing a special configuration of
generators in one orbit of arrangements, generating a given group $G_n$
with $n$ generators we find how can be added one reflection to obtain a bigger
group $G_{n+1}$. There are found ``universal'' sets of generators in each
group. These are very symmetric reducing the number of ways an additional
reflection can be added. They also allow to obtain representatives in all
orbits coming from the given group by simple permutations of the reflections
in them. Such universal graphs exist for all groups except $H_3,H_4,E_8$ and
$I_2(k),k=5$ or $k\ge 7$.
\item The extensions by one vertex of the universal graphs are studied.
For the groups without universal graphs are used sufficient samples of
quasi-universal graphs. It is shown that if the extension does not
contain a degenerate subgraph, in order to stay in a finite orbit
it must either be degenerate or represent generators in a
finite Coxeter group. For the remaining
extensions there is given a sequence of braid transformations which produce
a subgraph not belonging to a finite orbit.
For each degenerate extension it is demonstrated that in the minimal
realization (dim$(V)=$rank$(B)$) the extended graph represents a redundant set
of generators in a finite Coxeter group.
\item Using these results it is proved that every arrangement graph in a finite
orbit of the braid group represents, in its minimal realization, a set of
(possibly redundant) generators in a finite Coxeter group provided there is
not a number $k$ such that every subgraph with $k$ vertices to be degenerate,
but the rank of $B$ to be higher than $k$. It is shown that such property
is unstable under the action of the braid group which concludes the
classification of arrangements in finite orbits.
\item The orbits themselves are classified only in the extremal case of invertible
arrangement matrices. It is shown that the conjugacy class of the quasi-coxeter
element determines completely the orbit of the braid group for non-degenerate
configurations. The other extremal case of maximally singular rank=2 arrangement
matrices is treated in~\cite{Al} where additional invariants are introduced
in order to distinguish orbits with the same quasicoxeter element. In the
case of arrangement matrices of intermediate rank this work gives a criterion
of appurtenance to finite orbits.

\end{itemize}

\section{Orbits on the generators of Coxeter groups}

In this chapter are classified all orbits of the braid group action on non-redundant
generating reflections in finite Coxeter groups.
All the linear spaces will be supplied with
non-degenerate symmetric bilinear form for which the basis $\varepsilon_1,
\varepsilon_2\dotsc\varepsilon_n$ is orthonormal. This form gives a natural
isomorphism between $V$ and $V^*$ therefore each reflection can be given
by a nonzero vector $v$  $r_v(u)=u-2\frac{(v,u)}{(v,v)}v$. We will denote the
root systems $A_n,B_n,\dotsc$ and the corresponding Weyl groups by the same
letter and the meaning should be clear from the context. Sometimes, for
distinction, we will denote by $W(A_n),\dotsc$ the Weyl groups and extend
the same notation $W(H_{3,4})$ for the non-crystallographic Coxeter groups as well.

\subsection{Orbits on the generators of the classical\\
Coxeter groups $A_n,B_n,D_n$}

The root system $A_n=\{\varepsilon_i-\varepsilon_j,\,1\le i\neq j\le n+1\}$
generates a group $W(A_n)$ isomorphic to the symmetric group $S_{n+1}$ of
permutations of the basis vectors $\varepsilon_1\dotsc\varepsilon_{n+1}$. Reflections in
$W(A_n)$ correspond to transpositions in $S_{n+1}$. In this way the question
which reflections generate the group $W(A_n)$ transforms to the question which
transpositions generate the whole symmetric group $S_{n+1}$. We may represent
a set of $n$ transpositions on the $n+1$ basis vectors by a graph $\gamma$
with $n+1$ edges, numbered $1,2\dotsc n+1$ and edges corresponding to
transpositions $(i_k\ j_k)$. We call it the permutation graph in contrast
to the arrangement graph $\Gamma$.

\begin{lem}\label{th1}
The necessary and sufficient condition transpositions $(i_k,j_k)$,
$k=1\dotsc n$
to generate the whole symmetric group $S_{n+1}$ is not to exist two disjoint
proper subsets $A,B\subset\{1\dotsc n+1\}$ such that both $i_k,j_k$ to be either in
$A$ or in $B$ for all $k$.
This is equivalent to connectedness of the permutation graph
$\gamma$.
\end{lem}
\begin{proof}
The equivalence of the two conditions is immediate. If there exist two such
subsets then any product of transpositions will permute $A$ and $B$ without
mixing them. On the other hand if the graph is connected there
always exist a path joining any pair of vertices.
As $(i_1\ i_2)(i_2\ i_3)\dotsm(i_{l-1}\ i_l)=(i_1\ i_l)$
every transposition can be expressed by the given transpositions
hence they generate the whole symmetric group.
\end{proof}

A connected graph with $n+1$ vertices and $n$ edges is a tree. The numeration
of the vertices of this graph of transpositions is irrelevant to the
relative positions of these transpositions. Therefore we have a correspondence
between trees with numbered edges and the arrangements of reflections,
generating the group $W(A_n)$. 
\begin{lem}
The product of $n$ transpositions generating the group $S_{n+1}$ is a cycle
of length $n+1$.
\end{lem}
\begin{proof}
For $n=1$ the claim is trivial. Assume true for $n$. We have
\begin{equation}
(i_1\ j_1)(i_2\ j_2)\dotsm(i_n\ j_n)=(k_1\ k_2\dotsc k_{n+1}),\quad
\{k_1\dotsc k_{n+1}\}=\{1\dotsc n+1\}
\end{equation}
\begin{equation}
(k_1\ k_2\dotsc k_{n+1})(k_l\ n+2)=(k_1\dotsc k_{l-1}\ k_{n+2}\ k_{l+1}\dotsc
k_{n+1}\ k_l)
\end{equation}
\end{proof}

\begin{thm}\label{th2}
There is only one orbit of the braid group action on non-redundant sets of reflections
generating $W(A_n)$.
\end{thm}
\begin{proof}
We will show that a suitable braid transforms any set of generators to a
canonical one, whose permutation graph is linear, with edges numbered consecutively
$1,2,\dotsc,n$. For $n=1,2$ the claim is trivial.
Assume true for $n$. After ordering the first $n$ transpositions of the given
graph one obtains the graph in~Fig.\ref{fig2}.
\begin{figure}[htb]
\begin{center}\includegraphics{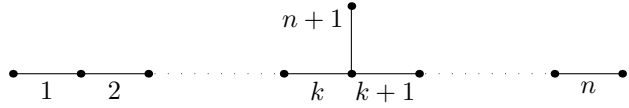}\end{center}
\caption{Induction hypothesis} \label{fig2}\end{figure}

Now acting with the braid $\sigma_{k+1}^2\sigma_{k+2}\dotsm \sigma_{n-1}\sigma_n$
is obtained a canonical graph with $n+1$ edges. Proof follows
by induction.
\end{proof}

Next group to be considered is $W(B_n)$. As it is known this is the group of
permutations and sign changes of the basis vectors in $n$-dimensional
Euclidean space so it is the semi-direct product $\Z_2^n\rtimes S_n$.
Each reflections corresponds to a sign change of one basis vector $\varepsilon_i
\mapsto -\varepsilon_i$, or a transposition of two basis vectors
$\varepsilon_i\leftrightarrow\varepsilon_j$ or a transposition with change of
sign $\varepsilon_i\leftrightarrow -\varepsilon_j$. These reflections
fall into two classes of conjugacy under the reflection group generated by
them. Call the transpositions with or without sign change the class $\cal A$
and the sign changes the class $\cal B$. These correspond to the long and short
roots in the root system $B_n$ or the opposite in $C_n$.

The group $W(B_n)$ acts transitively on the set of pairs $\{\varepsilon_i,
-\varepsilon_i\}$. The reflections of class $\cal B$ act trivially on this
set so the group $W(B_n)$ is generated by at least $n-1$ reflections of
class $\cal A$. The last generator must be in the other conjugacy class
$\cal B$.
We will present such a set of generators by a permutation graph with
$n$ vertices,
corresponding to the pairs $\{\varepsilon_i,-\varepsilon_i\}$ with $n-1$
numbered edges and one selected numbered vertex corresponding to the generator
of class $\cal B$. From Lemma\ref{th1} the graph should be a tree. Such graph
describes completely the relative positions of generating reflections.

\begin{thm}
The braid group action on generators of $W(B_n)$ has one orbit.
\end{thm}
\begin{proof}
Let the $k$-th generator be of class $\cal B$. Acting with
$\sigma_{n-1}\sigma_{n-2}\dotsm\sigma_k$ we change its number to $n$.
The remaining reflections
generate $W(A_n)$ and by Theorem~\ref{th2} there is a braid which brings them
to canonical configuration. The obtained graph is shown on~Fig.\ref{fig3}.
\begin{figure}[htb]
\begin{center}\includegraphics{fig.3}\end{center} \caption{} \label{fig3}
\end{figure}

A close inspection of the action \eqref{braid} on the graph convinces that
the braid $\sigma_{n-1}^2\dotsm\sigma_{k+2}^2\sigma_{k+1}^2\sigma_{k+2}\dotsm
\sigma_{n-2}\sigma_{n-1}$ transforms this graph to that of~Fig.\ref{fig4}
which is the canonical arrangement of reflections generating $W(B_n)$.
\begin{figure}[htb]
\begin{center}\includegraphics{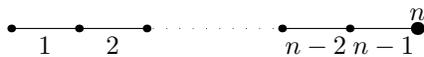}\end{center} \caption{Canonical permutation graph of $B_n$}
\label{fig4}
\end{figure}
\end{proof}

The last classical family of reflection groups is $W(D_n)$, $n=4,5,\dotsc$.
The group $W(D_n)$ acts on the $n$-dimensional Euclidean space by
permutations of the basis vectors and even number of sign changes. Reflections
form only one conjugacy class in the group and each of them
transpose two basis vectors with or without sign change
$\varepsilon_i\leftrightarrow\pm\varepsilon_j$. Again considering their
transitive action on the pairs $\{\varepsilon_i,-\varepsilon_i\}$ we see that
$n-1$ of them must generate the symmetric group, permuting these pairs. By
an isomorphism of the Euclidean space changing directions of the basis
vectors these reflections can be made transpositions without sign changes
of the basis vectors which forces the last reflection to be a transposition
with sign change.

Once more such arrangement will be presented by a permutation graph with
$n$ vertices corresponding to the pairs $\{\varepsilon_i,-\varepsilon_i\}$ and
indexed edges corresponding to the generating reflections. As $n-1$ edges
form a tree and there are $n$ vertices it follows that the graph is connected,
containing one cycle. Here one must allow two vertices of the graph to be
connected by two different edges, forming a cycle.

To count the orbits of the braid group on such arrangements we transform
them to a canonical form. As in the case of $W(B_n)$ we make the first $n-1$
to generate the symmetric group and order them to obtain a permutation graph
(Fig.\ref{fig5}).
\begin{figure}[htb]
\begin{center}\includegraphics{fig.5}\end{center} \caption{} \label{fig5}
\end{figure}

The product $r_1r_2\dotsm r_n$ which is invariant of the action of the braid
group seen as a permutation on the set of pairs $\{\varepsilon_i,
-\varepsilon_i\}$ decomposes into two cycles of lengths $k$, $n-k$ so
there are at least $\lfloor \frac{n}{2}\rfloor$ orbits. The graph
of Fig.\ref{fig5} is transformed to a similar one with $k'=n-k$ by the braid
$$\sigma_{n-k-1}^{-1}\sigma_{n-k}\dotsm\sigma_{n-2}\dotsm\sigma_2^{-1}\sigma_3
\dotsm\sigma_{k+1}\sigma_1^{-1}\sigma_2\dotsm\sigma_{k-1}\sigma_k\sigma_1\dotsm
\sigma_{k-2}\sigma_{k-1}$$
and so there are exactly
$\lfloor \frac{n}{2}\rfloor$ orbits of the braid group acting on $n$-tuples
of reflections generating $W(D_n)$.

\begin{figure}[htb]
\begin{center}\includegraphics{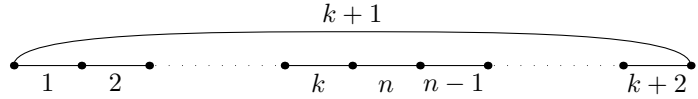}\end{center} \caption{Universal graph for $D_n$}
\label{fig6}
\end{figure}

The graph on Fig.\ref{fig6} can also be used as canonical for the orbits on
generators of $W(D_n)$.
It is unique for each orbit and universal in the sense that all the orbits are
obtained by different numberings of its edges.

The graph of transpositions is unique for every arrangement except for
the generators of $W(D_4)$, where many isomorphisms appear. 
\begin{figure}[htb]
\begin{center}\includegraphics{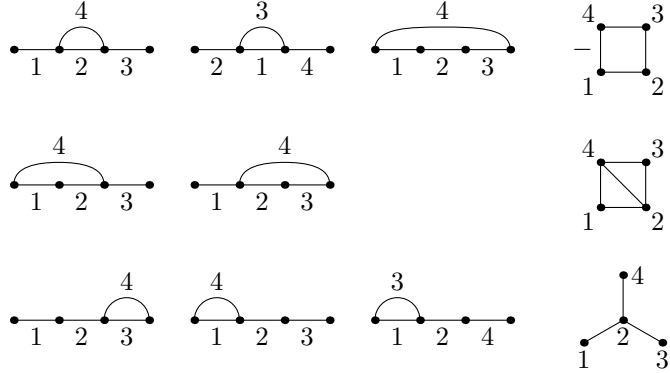}\end{center}
\caption{Permutation graphs, corresponding to the same arrangement}
\label{fig7}
\end{figure}
The equivalent transposition graphs of these generators are shown in the rows
of Fig.\ref{fig7}, while in the last column are shown the arrangement graphs,
corresponding to them. Written explicitly it is
easy to see that these isomorphisms are reflections in the Euclidean space.
If these reflections are added to the group $W(D_4)$ one obtains the group
$W(B_4)$. In fact our presentation with graphs of transpositions, which do not give information
whether these transpositions of basis vectors include sign changes or not, is
loose enough to hide all the isomorphisms, which if added to $W(D_n)$ give
the group $W(B_n)$. In the case of $W(D_4)$ the outer isomorphisms coming
from Fig.\ref{fig7} and those from sign changes of the basis vectors are
independent and when both added the group obtained is $W(F_4)$.

\subsection{Orbits on the generators of exceptional Coxeter groups}

There is no obvious interpretation of these groups as permutation groups.
Because of that, we will use the arrangement graphs. As it is seen from
the definition~\eqref{braid} the action of $i$-th elementary braid
coincides with the result of conjugation of the $i+1$-th reflection by
$i$-th $r_{i+1}\mapsto r_ir_{i+1}r_i$ followed by their transposition
$r_i\leftrightarrow r_{i+1}$. In terms of the graph, conjugation of the 
reflection corresponding to $i+1$-th vertex will affect only edges incident
with this vertex. Moreover, the resulting edge $g(i+1,k)$ will depend
only on $g(i,i+1)$, $g(i+1,k)$, and $g(i,k)$. This dependence is given
in Table~A.

New arrangements are built inductively by adding one vertex to an
arrangement which is known to be in finite orbit. In order to
minimize the possibilities of such extensions it is convenient to pick up
the most uniform
arrangement in every orbit. As a byproduct these uniform arrangements
are also universal i.e. by changing only the ordering of their
vertices are obtained arrangements in all orbits of the braid group
on generators of a given group.

The braids
\begin{align}
\label{trans1}
\sigma_i(r_1,r_2,\dotsc,r_n)=(r_1,\dotsc,r_{i+1},r_i,\dotsc,r_n)\qquad
{\rm if}\quad r_ir_{i+1}=r_{i+1}r_i\\
\label{trans2}
\sigma_{n-1}\sigma_{n-2}\dotsm\sigma_1(r_1,r_2,\dotsc,r_n)
=(r_1r_2r_1,r_1r_3r_1,\dotsc,r_1r_nr_1,r_1r_1r_1)
\end{align}
preserve the graph, which represents only relative positions
of the reflections; but they affect the ordering. We will find which orderings
are obtained  by these transformations.

\begin{lem}\label{ord1}
All permutations of the vertices of an arrangement graph, which is a tree
are obtained by the action of the braid group.
\end{lem}
\begin{proof}
Any numbering can be transformed by the braids~\eqref{trans1},\eqref{trans2}
to a fixed one with the property that the first vertex is a leaf in the tree,
and the induced sub-graphs on vertices $1,2,\dotsc,k$ for $k=1,2,
\dotsc,n$ are trees. Arbitrary numbering is defined by a permutation
$i:\{1,\dotsc,n\}/to\{1,\dotsc,n\}$ of the vertices.
By a cyclic permutation from the braid~\eqref{trans2} it is always possible
to make $i(1)=1$. Assume the vertices to be ordered up to the number $k$ and
$i(k+1)=p>k+1$. If $i^{-1}(p)$ and $i^{-1}(p-1)$ are not joined, $\sigma_{p-1}$
lowers the index of $k+1$. If they are joined lower first the $i^{-1}(p-1)$.
This process will continue until either $i(k+1)=k+1$ or $i(k+1)=p$ and
$i^{-1}(p),i^{-1}(p-1),\dotsc,i^{-1}(k+1)$ form a path in the tree.
In the last case the braid $\sigma_{k-1}\sigma_{k-2}\dotsm\sigma_1$ effectively
rises the numbers of the first $k$ vertices and conjugates the remaining ones
with $r_1$, the last being trivial if $k\ge 2$ as $1$ is joined only with $2$
by assumption on the fixed numbering.
After cyclic permutation the first $k$ numbers are restored and the path obtains
the numbers $p-1,p-2,\dotsc,k+1,n$. 
If this happens with $k=1$, the braid
$(\sigma_{n-1}\sigma_{n-2}\dotsb\sigma_{2})^{-1}$ preserves the graph and the path
obtains the numbers $p-1,p-2,\dotsc,2,n$.
It is always possible to lower the index of $i(k+1)>k+1$ eventually bringing
it to $i(k+1)=k+1$. By induction all the vertices can be ordered
so that $\forall k, i(k)=k$.
\end{proof}

\begin{lem}
There exist $n-1$ orbits of the transformations
\eqref{trans1}-\eqref{trans2} of a graph with $n$ vertices, which is a cycle.
\end{lem}
\begin{proof}
Fix a linear ordering of the vertices such that $k,k+1\mod n$ to be joined
for all $k\in\Z_n$. Arbitrary ordering will be denoted by $i(k)$.
The quantities
\begin{equation}\label{quant}\begin{split}
q_<&=\#\{k\in\Z_n, i(k)<i(k+1\mod n)\}\\
q_>&=\#\{k\in\Z_n, i(k)>i(k+1\mod n)\}
\end{split}\end{equation}
are invariants of the transformations \eqref{trans1}-\eqref{trans2}.
We may order the vertices so that $i(k)=k$ for $k\leq q_<$ in the manner of the
previous lemma hence these invariants are the only obstruction.
As $q_<+q_>=n$, $1\le q_<\le n-1$ there are $n-1$ orbits.
\end{proof}

The quantities \eqref{quant} are preserved also when the cycle is part of the
graph. Such invariants, associated to each directed cycle, will classify the
orbits of \eqref{trans1}-\eqref{trans2}. These invariants are not independent.
If ${\cal C}={\cal C}^1+{\cal C}^2$ in the first homology group, the quantities
$(q_<,q_>)$ associated to ${\cal C}$ are expressed through
$(q_<^{1,2},q_>^{1,2})$, associated to ${{\cal C}^{1,2}}$
\begin{equation}
(q_<,q_>)=(q_<^1+q_<^2-k,q_>^1+q_>^2-k)
\end{equation}
where $k$ denotes the number of edges, common to ${\cal C}^1$ and ${\cal C}^2$.

Let directed cycles ${\cal C}^1,\dotsc,{\cal C}^k$ form a basis in the first
homology group of an unindexed graph and $l({\cal C}^i)$ be the length of the cycle
${\cal C}^i$. This graph generates at most $(l({\cal C}^1)-1)
(l({\cal C}^2)-1)\dotsm(l({\cal C}^k)-1)$ orbits. However,
due to symmetries between these cycles and restrictions on some of the
invariants by fixing the others the number of orbits is usually much lower.

\begin{lem}\label{ord2}
A graph $\Gamma$ can be transformed by \eqref{trans1}-\eqref{trans2} to a
similar one with consecutive indices on a pair of vertices
$A$ and $B$ if and only if there is not a cycle in which $A$ and $B$ are not
neighboring, and with one of its invariants $q_<,q_>$ equal to $1$.
\end{lem}
\begin{proof}
We proceed as in Lemma~\ref{ord1}. Let the vertices are labeled by letters
and indexed by
$$i:\{A,B,\dotsc\}\to\{1,2,\dotsc,n\}.$$
For easier notation the operations on indices are done in $\Z_n$.
 The index of $B$ is lowered by
$\sigma_{i(B)-1}$ if $i^{-1}(i(B)-1)$ is not connected to $B$. If
$B,i^{-1}(i(B)-1),i^{-1}(i(B)-2),\dotsc,i^{-1}(i(B)-l)$ is a path
$\sigma_{i(B)-l-1}$ shortens it. If $i^{-1}(i(B)-l)=A$, the index of $B$
cannot be lowered, and there is a cycle $A,B,C,\dotsc,D$ with one rising of
the index. In such a case we repeat the procedure with rising the index of $B$.
Eventually, either $i(B)=i(A)-1$ or there is a cycle $B,A,E,\dotsc,F$ with one
rising of the index. In the second case the cycle $A,E,\dotsc,F,B,C,\dotsc,D$
has one of the invariants equal to $1$ and the transformations
\eqref{trans1}-\eqref{trans2} alone cannot make $A$ and $B$ with consecutive
indices.
\end{proof}

\subsection{Orbits on generators of the groups $E_6,E_7,E_8$}

The most symmetric arrangement in the orbit of generators of $W(A_n)$ is
the complete graph $\Gamma_0(A_n)$ corresponding to the matrix
\begin{equation}
B(A_n)=\begin{pmatrix}2&1&\cdots&1\\
1&2& &1\\
\vdots& & \ddots &\vdots\\
1&\hdotsfor{2}&2
\end{pmatrix}.
\end{equation}
Indeed its symmetry group is the group of all permutations of the vertices
which is much bigger than the group $\Z_2$ of symmetries of the Dynkin diagram
of $W(A_n)$.

An extension of this graph by one vertex and edges labeled $\pm3$ is
determined by the number $k$ of these edges, and the difference between
the number of positive and negative labels. Postponing consideration of
extensions of configurations with degenerate matrices  we
see that all edges must have equal sign which can be taken positive.

The arrangement matrix of one vertex extension of $A_n$ is
\begin{equation}
B(A_n,k)=\begin{pmatrix}
2&1&\hdotsfor{2}&1&0\\
1&2&1&\cdots&1&\vdots\\
\vdots&&\ddots&&\vdots&1\\
\vdots&&&\ddots&1&\vdots\\
1&\hdotsfor{2}&1&2&1\\
0&\cdots&1&\cdots&1&2
\end{pmatrix},
\end{equation}
where on the last row and column there are $k$ $1$s. The determinant of this
matrix is calculated using
\begin{equation}\label{ind1}
\det(B(A_n))=\begin{vmatrix}
2&1&\cdots&1\\
1&\ddots&&\vdots\\
\vdots&&2&1\\
1&\cdots&1&2
\end{vmatrix}=n+1\qquad
\begin{vmatrix}
2&1&\cdots&1\\
1&\ddots&&\vdots\\
\vdots&&2&1\\
1&\cdots&1&1
\end{vmatrix}=1
\end{equation}

\begin{equation}\label{detD}
\det(B(A_n,k))=2(n+1)-k(n-k+1).
\end{equation}

Identities \eqref{ind1} are proved by induction. The requirement of
non-de\-gene\-racy reads $2(n+1)-k(n-k+1)\ne 0$. Solving for $n$ the opposite
condition
\begin{equation}
n=\frac{k^2-k+2}{k-2}\qquad
\begin{array}{lccccc}
k=&0&1&2&3&4\\
n=&-1&-2&\infty&8&7
\end{array}.
\end{equation}

One sees that $\det(B(A_n,k))\ne 0$ for any $n$ if $k=1,2,n-1,n$. It is also
satisfied for $k=3,5,6$ if $n<8$ and $k=4,n<7$. Any extension out of these
restrictions will contain a degenerate sub-graph and is not considered here.
The extensions by $k=1,n$ give an $A_{n+1}$ arrangement, those with
$k=2,n-1$ give a $D_{n+1}$ and $k=3,\,5<n<8$;
$k=4,\,n=6$; $k=5,\,n=7$ give an $E_{n+1}$ arrangement.

We continue with the extensions of $D_n$ arrangements. The most uniform and
universal $D_n$ arrangement is the extension of the symmetric arrangement
of $A_n$ by $n-1$ edges. It will be denoted $\Gamma_0(D_n)$ and corresponds
to a complete graph with one edge
deleted. If the ends of this edge are $v_a,v_b$ and the vertices are indexed
$i:V\to\{1,2,\cdots,n\}$ according to their ordering, the difference
$|i(v_a)-i(v_b)|\mod n$ determines the different orbits of the braid group.

Let $\Gamma_0(D_n)$ be extended to $\Gamma'$ with the vertex $v'$.
The non-degeneracy does not depend on the order of vertices therefore
the most general extension of $\Gamma_0(D_n)$ is one of the following:

\begin{enumerate}
\item Extension with $k$ edges for which $\{v_a,v'\},\{v_b,v'\}\not\in E'$

\begin{equation}\label{detE1}
\det(B(D_n,k)_1)=
\begin{vmatrix}
2&1&\hdotsfor{2} & 1&0&0\\
1&2&1&\cdots & 1&1&\vdots\\
\vdots& &\ddots & & &\vdots&1\\
\vdots& & &\ddots & &\vdots&\vdots\\
1&\hdotsfor{2} & &\ddots &1&1\\
0&1&\hdotsfor{2} & 1&2&0\\
0&\cdots&1 &\cdots & 1&0&2
\end{vmatrix}=8-4k.
\end{equation}

It is non-degenerate only for $k=1$ and the obtained arrangement is $D_{n+1}$.

\item Extension with $k+1$ edges for which
$\{v_a,v'\}\in E',\,\{v_b,v'\}\not\in E'$
\begin{equation}\label{detE2}
\det(B(D_n,k)_2)=
\begin{vmatrix}
2&1&\hdotsfor{2} & 1&0&0\\
1&2&1&\cdots & 1&1&\vdots\\
\vdots& &\ddots & & &\vdots&1\\
\vdots& & &\ddots & &\vdots&\vdots\\
1&\hdotsfor{2} & &\ddots &1&1\\
0&1&\hdotsfor{2} & 1&2&1\\
0&\cdots&1 &\cdots & 1&1&2
\end{vmatrix}=8-n.
\end{equation}

It is non-degenerate for $n<8$ and the obtained arrangement is $E_{n+1}$ if
$n\ge 5$, $D_5$ if $n=4$, and $A_4$ if $n=3$. Call such arrangement
$\Gamma(D_{n-1}\subset E_n,k)$.

\item Extension with $k+2$ edges for which
$\{v_a,v'\},\{v_b,v'\}\in E'$.
Up to now, in order to avoid degenerate sub-graphs, it was assumed that
all the new edges had positive signs . Here it is possible
only for $k=n-2$. If $0<k<n-2$ there exist vertices $v_c,v_d$ such that
$\{v_c,v'\}\in E'$, $\{v_d,v'\}\not\in E'$. The subgraph on vertices
$v_a,v_d,v_b,v_v'$ is a cycle as are the triangles $v_a,v_c,v'$, and
$v_b,v_c,v'$. Non-degenerate cycles have odd number of
negative signed edges. It is easy to see that the three cycles cannot 
be simultaneously non-degenerate.
The only permitted extensions are those with $k=0,n-2$ giving in both cases $D_{n+1}$.

\end{enumerate}

By Lemma~\ref{ord2} the extension $A_n\subset E_{n+1}$ can always be braid
transformed to make the indices of the ends of one of the new
edges consecutive say $i,i+1$. Applying the appropriate braid $\sigma_i$ or
$\sigma_i^{-1}$ it is transformed to a graph
$\Gamma(D_n\subset E_{n+1},n-1)$. For $E_6$ this is actually an
universal graph. We will prove that $\Gamma(D_6\subset E_7,2)$ is universal
for $E_7$ i.e. all extensions $\Gamma(D_6\subset E_7,k)$, $\Gamma(E_6\subset
E_7)$ and $\Gamma(A_6\subset E_7)$ can be braid transformed to it. The group
$E_8$ does not have an universal graph so we will use
of two unordered graphs of its generators with the property that every orbit
contains at least one of them.

Every connected subgraph of $\Gamma(E_6)$ with 5 vertices is either
$\Gamma(A_5)$ or $\Gamma(D_5)$. Our approach to finding the orbits of the
braid group on arrangements generating $W(E_6)$ will be to consider the graphs
$\Gamma(D_5\subset E_6)$, in which the first 5 vertices belong to the
subgraph $\Gamma(D_5)$. The braid
$$\tau_4^5=(\sigma_4\sigma_3\sigma_2\sigma_1)^5$$
preserves the subgraph $\Gamma(D_5)$ and conjugates the last reflection
$$r_6\mapsto r_1r_2r_3r_4r_5r_6r_5r_4r_3r_2r_1.$$
To identify different $E_6$ orbits and arrangements $\Gamma(D_5\subset E_6)$
in them one may use the following procedure. First are listed all
extensions of $\Gamma_0(D_5^{(1)})$ and $\Gamma_0(D_5^{(2)})$, and grouped
into sets of transitive action of the braid $\tau^5$. Then, in each member of
these sets are considered other $D_5$ subgraphs. Appropriate braid will make
the indices of these subgraphs to take the values $1,\dotsc,5$ obtaining a new
graph in the given class. When these new graphs fall in different sets, the sets
are unified. At the end one obtains a list of sets of $\Gamma(D_5\subset E_6)$
graphs representing different orbits. The result is that the graph
$\Gamma_0(E_6)=\Gamma(D_5\subset E_6,4)$
is universal for $W(E_6)$ where the orbit depend on the indices $i,j,k$. This
graph is symmetric with respect to $j,k$ but actually any permutation of
the indices $i,j,k$ yields a graph in the same orbit. Using the fact that the
braid $\tau_6=\sigma_5\sigma_4\dotsm\sigma_1$ permutes cyclically the indices
we see that the orbit depend on the relative positions
of $i,j,k$ in $\Z_6$ or in other words the orbits correspond to different
inscribed triangles in the regular hexagon (Fig.\ref{fig10}).
\begin{figure}[htb]
\hfil\begin{minipage}[t]{.24\textwidth}
\includegraphics{fig.9}
\caption{$\Gamma_0(E_6)$} \label{fig9}
\end{minipage}
\begin{minipage}[t]{.4\textwidth}
\includegraphics{fig.10}
\caption{} \label{fig10}
\end{minipage}\hfil
\end{figure}

The subgraphs of arrangements generating $W(E_7)$ are
$A_6,D_6^{(k)},E_6^{(k)}$. A detailed inspection shows that all extensions of
$\Gamma_0(E_6)$ to $\Gamma(E_7)$ contain subgraphs generating $W(D_6)$ which
allows us to proceed in the same way as with $E_6$. There are four orbits
coming from different inscribed triangles in the regular heptagon
(Fig.\ref{fig11}) and one more orbit which does not have graph
$\Gamma(D_6\subset E_7,5)$. One graph $\Gamma(D_6\subset E_7,4)$ in the last
orbit is shown on Fig.\ref{fig12}, where the three vertices in the center,
with respect to which the graph is symmetric, have indices $1,3,6$.
\begin{figure}[htb]
\begin{minipage}[t]{.75\textwidth}
\includegraphics{fig.11}
\caption{} \label{fig11}
\end{minipage}\hfil
\begin{minipage}[t]{.25\textwidth}
\includegraphics{fig.12}
\caption{$E_7^{(5)}$} \label{fig12}
\end{minipage}
\end{figure}

The last group $W(E_8)$ can be generated by a reflection
configuration, in which all subgraphs with 7 vertices generate $W(E_7)$. Aside
from that, there are 5 orbits coming from the graph $\Gamma(D_7\subset E_8,6)$
with indexing of the vertices corresponding to the 5 different inscribed
triangles in the regular octagon (Fig.\ref{fig13}).
There are also 3 orbits which have graph $\Gamma(D_7\subset E_8,5)$,
and one more orbit in which all graphs contain
only $\Gamma(E_7)$ subgraphs. A very symmetric representative in the last one
is shown on Fig.\ref{fig14}.
\begin{figure}[htb]
\begin{minipage}[t]{.55\textwidth}
\includegraphics{fig.13}
\caption{} \label{fig13}
\end{minipage}\hfil
\begin{minipage}[t]{.25\textwidth}
\includegraphics{fig.14}
\caption{$E_8^{(9)}$} \label{fig14}
\end{minipage}
\end{figure}

As we have seen, the quasicoxeter element is invariant under the action of
the braid group. It can be computed for every reflection arrangement
using~\eqref{quasicox1}. The eigenvalues of quasicoxeter elements of
reflection arrangements, generating finite Coxeter groups must be roots of
unity, moreover, in case of simply-laced groups $A_n,D_n,E_6,E_7,E_8$ the
characteristic polynomial factors into cyclotomic polynomials.
Recall that the $n$-th cyclotomic polynomial is given by
\begin{equation}
\Phi_n(x)=\prod_{\substack{1\le k\le n\\ \gcd(k,n)=1}}
(x-e^\frac{2\pi\imath k}{n})\,.
\end{equation}
The characteristic polynomials of quasicoxeter elements corresponding to the
different orbits of arrangements, generating $W(E_n)$ up to constant
factors are
\[
\begin{array}{cc}
\mbox{orbit} & \det(C-\Id x)\\ \hline
E_6^{(1)}& \Phi_3(x)\Phi_{12}(x)\\
E_6^{(2)}& \Phi_9(x)\\
E_6^{(3)}& \Phi_3(x)\Phi_6(x)^2\\
E_7^{(1)}& \Phi_2(x)\Phi_{14}(x)\\
E_7^{(2)}& \Phi_2(x)\Phi_6(x)\Phi_{12}(x)\\
E_7^{(3)}& \Phi_2(x)\Phi_{18}(x)\\
E_7^{(4)}& \Phi_2(x)\Phi_6(x)\Phi_{10}(x)\\
E_7^{(5)}& \Phi_2(x)\Phi_6(x)^3\\
E_8^{(1)}& \Phi_{30}(x)\\
E_8^{(2)}& \Phi_{24}(x)\\
E_8^{(3)}& \Phi_{20}(x)\\
E_8^{(4)}& \Phi_{6}(x)\Phi_{18}(x)\\
E_8^{(5)}& \Phi_{15}(x)\\
E_8^{(6)}& \Phi_{12}(x)^2\\
E_8^{(7)}& \Phi_{10}(x)^2\\
E_8^{(8)}& \Phi_{6}(x)^2\Phi_{12}(x)\\
E_8^{(9)}& \Phi_{6}(x)^4\\
\end{array}
\]

\subsection{Orbits on generators of the groups $F_4,H_4$}

\begin{figure}[htb]
\begin{center}\includegraphics{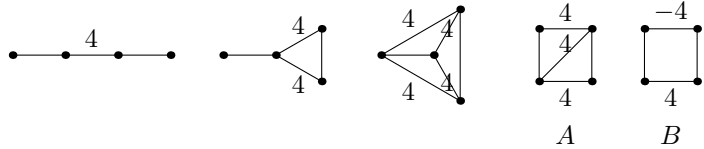}\end{center}
\caption{The two orbits of $F_4$} \label{fig8}
\end{figure}

In view of their shortness these orbits can be computed manually using
Table~A and Lemma~\ref{ord2}. Canonical generators of $F_4$ lie
in the orbit with graphs on Fig.\ref{fig8},
where the invariants \eqref{quant} of the squares in graphs $A,B$ are
$(q_<,q_>)=(2,2)$ and $(1,3)$ correspondingly.
The graphs $A$ and $B$ with invariants $(1,3)$ and $(2,2)$ form
another orbit of the braid group and these two orbits contain all the
arrangements of reflections, generating $W(F_4)$.

For all crystallographic Coxeter groups, considered up to here, there
was a uniform procedure for finding all the configurations, generating a given
group. Starting with the canonical generators and acting by the braid group
and permuting the indices, it was possible to obtain all configurations
generating a given group.
This procedure fails for any group which has an arrangement of generating
reflections involving matrix elements $2\cos\frac{k\pi}{n}$, $n=5,n\ge 7$.
The reason is that for an abstract group with two generating reflections
$r_1,r_2$, $r_1^2=r_2^2=(r_1r_2)^n=\Id$ there exist more than one
linearly non-isomorphic realizations if $n=5$ or $n\ge 7$. More precisely,
the number of such realizations is equal to the number of regular star-polygons
with $n$ sides plus one, or the whole part of half the Euler's totient function
$\lfloor\frac{\phi(n)}{2}\rfloor$.

In order to obtain all arrangements generating $W(H_4)$, one must
allow transformations
\begin{align}
\label{startr1}(r_i,r_j)&\mapsto (r_j,r_i)\\
\label{startr2}(r_i,r_j)&\mapsto (r_i,r_ir_jr_i)\\
\label{startr}(r_i,r_j)&\mapsto(r_i,r_jr_ir_j)\quad\mbox{if }(r_ir_j)^5=\Id\,.
\end{align}

These arrangements split into families such that arrangements from the same
family are obtained by transformations not involving~\eqref{startr}.
Triples of generating reflections of the group $H_3$ form 3 families containing
one orbit each. In every orbit there is a linear graph corresponding to a pair
of reciprocal regular polyhedra or star-polyhedra of Kepler-Poinsot~\cite{Cox1}.

The group $W(H_4)$ has five families of generating arrangements and
in each family there is at least one arrangement, whose graph is linear.
These linear graphs correspond to pairs of reciprocal regular
star-polyhedra in the four-dimensional space.
In each family of arrangements there are two or three orbits.

The list of orbits according to their family is given in Table~B. There are given
also the characteristic polynomials of quasicoxeter elements in each
orbit. Notice that the transformations \eqref{startr1}-\eqref{startr2}
preserve $\det(B)$, while~\eqref{startr} does not, therefore the families can be
characterized by $\det(B)$:
\begin{equation}
\begin{array}{lccccc}
{\rm family}& H^A&H^B&H^C&H^D&H^E\\
\det(B)&\frac{7-3\sqrt{5}}{2}&\frac{7+3\sqrt{5}}{2}&\frac{3+\sqrt{5}}{2}&
\frac{3-\sqrt{5}}{2}&1
\end{array}
\end{equation}

We conclude with the remark that in each family of orbits there are
universal graphs:
\begin{figure}[htb]
\begin{center}\includegraphics{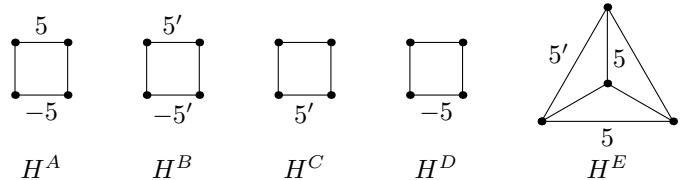}\end{center}
\caption{Universal graphs for the families of orbits of $H_4$}
\label{figh24}
\end{figure}

\section{Extensions by one vertex of the universal graphs}

Here will be considered arrangements in which every subarragement generates
finite Coxeter group. According to \cite{DM} and
Lemma~\ref{subgraph} only such configurations may belong to finite orbits of
the braid group. As it is always possible to bring the subconfiguration to
its universal graph we will consider only extensions of the universal
graphs. By so doing, the task is simplified in two ways: only one graph is
considered for all orbits of the braid group on configurations generating
particular Coxeter group; and the universal graphs are deliberately chosen
to have big symmetry groups to reduce the number of possible extensions.

\begin{defn}
An {\it admissible extension} of a graph $\Gamma$ is an extension by one vertex,
such that the obtained graph does not contain degenerate subgraphs, nor
subgraphs with infinite orbit under the braid group action.
\end{defn}

Only extensions which do not contain degenerate subarrangements are
considered. The remaining extensions are treated in the next section.
When talking about realization of a degenerate arrangement matrix it is
always understood the minimal realization ${\rm rank}(B)={\rm dim}(V)$ as
only in this case the arrangements have a simple meaning of redundant generators in
finite Coxeter group. As a demonstration for redundancy will be
given expressions for one of the reflections through the others.

\subsection{Extensions of $H_3$, $H_4$ configurations}

\begin{figure}[htb]
\begin{center}\includegraphics{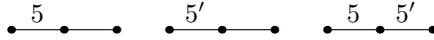}\end{center}
\caption{Representatives in the orbits of $H_3$} \label{figh12}
\end{figure}

First we consider extensions of the graphs on Fig.\ref{figh12},
which are arrangements in the three orbits of ${\cal B}_3$, generating $W(H_3)$.

All degenerate admissible extensions of the graphs in Fig.\ref{figh12} are
shown in fig.~\ref{figh14}. They represent redundant generators of $W(H_3)$.
The explicit expressions of one of the reflections
through the others is given in Table~C.
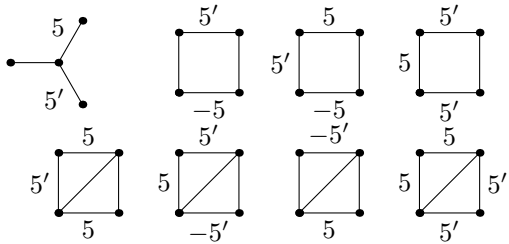
\begin{figure}[htb]
\begin{center}\includegraphics{figh.13}\end{center}
\end{figure}\vskip-3ex
\begin{figure}[htb]
\begin{center}\includegraphics{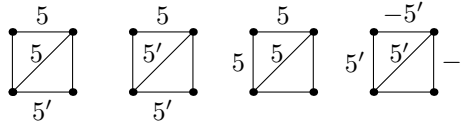}\end{center}\caption{Degenerate extensions} \label{figh14}
\end{figure}

\begin{figure}[htb]
\begin{center}\includegraphics{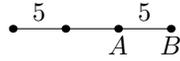}\end{center}
\caption{Admissible extension belonging to an infinite orbit} \label{figh15}
\end{figure}
The extensions, which do not generate $W(H_4)$ e.g. Fig.\ref{figh15}
can always be transformed by braids~\eqref{trans1}-\eqref{trans2}
according to Lemma~\ref{ord2} to a new indexing of the vertices, in which
$A,B$ have indices $1,2$. Then the braid $\sigma_1$ if $A=1,B=2$ or
$\sigma_1^{-1}$ if $A=2,B=1$ transforms the graph to a new one with
a subgraph not generating finite three-dimensional Coxeter group
and according to~\cite{DM} does not stay in a finite orbit of the braid
group. The same argument applies to all non-degenerate admissible extensions
of the three graphs in Fig.\ref{figh12}, which are enlisted in table~D
together with determinants of the arrangement matrices.
All the remaining admissible extensions represent arrangements of reflections,
generating $W(H_4)$.

\begin{figure}[htb]
\begin{center}\includegraphics{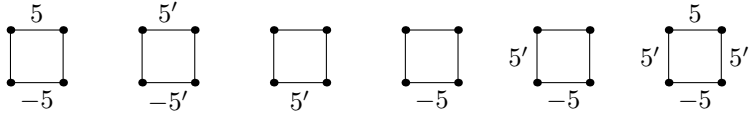}\end{center}
\caption{uniformized universal graphs for $H_4$}\label{figh19}
\end{figure}
\begin{figure}[htb]
\begin{center}\includegraphics{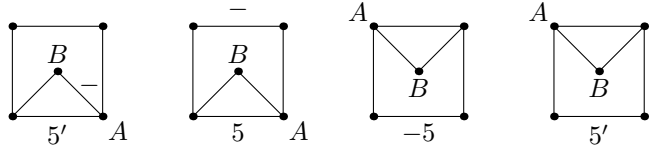}\end{center}
\caption{Admissible extensions belonging to infinite orbits}\label{figh20}
\end{figure}
For generators of the group $W(H_4)$ one may use the universal graphs
on Fig.\ref{figh19}.
There are two graphs for the last family of orbits for uniformity. Most of the
extensions of these graphs by one vertex, in which all subgraphs represent
non redundant generators of finite Coxeter groups are degenerate. These
degenerate extensions are given in Table~E, with explicit formulas for
one of the reflections through others. In order to save space
the numbering of the vertices is not given in the table. The convention is
to index the vertices counterclockwise beginning with the upper left vertex;
the central vertex has index~5.
Apart from that, the non-degenerate admissible extensions are shown
in Fig.\ref{figh20}.
All these graphs, when transformed in a way analogous to that of extensions
of $H_3$ obtain subgraphs, which do not belong to finite orbits.

\subsection{Extensions of the universal graphs of the Weyl groups}

There is no need to consider admissible extensions with edges labelled $\pm5,5'$
as they are also extensions of $H_3$ or $H_4$ arrangements.

\subsubsection{Extensions of $B_n,F_4$}
\begin{figure}[htb]
\begin{center}\includegraphics{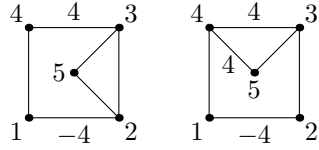}\end{center}
\caption{Admissible extensions of $\Gamma_0(F_4)$} \label{figh22}
\end{figure}

The admissible extensions of the universal arrangement of $W(F_4)$ are only two
(Fig.\ref{figh22}). They are degenerate. The reflection
corresponding to the fifth vertex is equal to $r_3r_4r_1r_2r_1r_4r_3$ for the
first extension and to $r_4r_1r_2r_1r_4$ for the second one.

The extensions of $\Gamma_0(B_n)$, in which all subgraphs with three
vertices are non-degenerate and have finite orbits, fall in the following
three cases:
\begin{equation}
\begin{pmatrix}B_1&b_1&a_1\\
b_1^t&2&0\\
a_1^t&0&2\end{pmatrix},\
\begin{pmatrix}B_1&b_1&b_2\\
b_1^t&2&1\\
b_2^t&1&2\end{pmatrix},\
\begin{pmatrix}B_1&b_1&a_2\\
b_1^t&2&\sqrt2\\
a_2^t&\sqrt2&2\end{pmatrix}
\end{equation}
where the submatrices $B_1,b_1,a_1,b_2,a_2$ are
\begin{gather}
B_1=\begin{pmatrix}2&1&\hdotsfor{2}&1\\
1&2&1&\cdots&\vdots\\
\vdots&\ddots&\ddots&\ddots&\vdots\\
\vdots&&\ddots&\ddots&1\\
1&\hdotsfor{2}&1&2\end{pmatrix},\quad
\begin{array}{c}b_1^t=(\overset{n}{\overbrace{\sqrt{2},\dotsc,\sqrt{2}}})\\
b_2^t=(\overset{p}{\overbrace{\sqrt{2},\dotsc,\sqrt{2}}},0,\dotsc,0)
\end{array}\\
\label{a1}
a_1^t=(0,\dotsc,0,\overset{q}{\overbrace{-1,\dotsc,-1}},
\overset{p}{\overbrace{1,\dotsc,1}})\,,\quad
a_2^t=(1,1,\dotsc,1)\,.
\end{gather}

Determinants of the matrices of the three extensions respectively, are
\begin{equation}
2(2-p-q)\,,\quad 4-n\,,\quad 2
\end{equation}
The first extension is degenerate for $p=2,q=0$ or $p=0,q=2$, or $p=1,q=1$.
The first two possibilities coincide as arrangements using the
identification~\eqref{freed}. They are realized by $r_{n+1}=
r_{n-1}r_nr_{n-2}r_nr_{n-1}$, while the third -- by $r_{n+1}=
r_{n-2}r_{n-1}r_{n-2}$. The only non-degenerate extensions without degenerate
subgraphs are for $p=1,q=0$ or $p=0,q=1$ which are equivalent. In this
case the extended arrangement generates $B_{n+1}$.

The extension of the second case is degenerate for $n=4$. It can be realized by
\begin{equation}
r_2=g_1rg_2r_1g_2^{-1}rg_1^{-1}
\end{equation}
where $g_1=r_4r_1r_3r_1$, $g_2=r_1r_4$, and $r$ is a reflection depending
on the number $p$:
\begin{equation}
\begin{array}{rlrl}
r&=r_5&\mbox{if}&p=0\\
r&=r_4r_1r_5r_1r_4&\mbox{if}&p=1\\
r&=r_4r_1r_4r_5r_4r_1r_4&\mbox{if}&p=2\\
r&=r_4r_5r_4&\mbox{if}&p=3
\end{array}
\end{equation}
These are redundant generators in the group $F_4$ which explains why we
have expressed $r_2$ instead of $r_5$. The non-degenerate extensions when
$n=1,2,3$ generate the groups $B_2,B_3,F_4$ correspondingly.

The third case is non-degenerate and by exchanging the last two
reflections it becomes $\Gamma_0(B_{n+1})$.

\subsubsection{Extensions of the simply-laced graphs}

No admissible extensions with edges labelled $\pm5,5',4$ need to be considered
as these are also extensions of already examined graphs.
We begin with the extensions of $\Gamma_0(A_n)$.
\begin{equation}
\det\begin{pmatrix}B_1&a_1\\ a_1^t&2\end{pmatrix}
=(p-q)^2-(p+q)(n+1)+2(n+1),
\end{equation}
where $a_1$ is the column vector~\eqref{a1}.
It is convenient to assume $p>q$ as the expression is symmetric with
respect to $p$ and $q$.
To analyze when this determinant vanishes it is convenient to introduce new
variables
\begin{equation}
\begin{split}p-q&=u\\p+q&=v\,.\end{split}
\end{equation}

Solving for $v$
\begin{equation}
v-2=\frac{u^2}{n+1}\,.
\end{equation}

We are looking for solutions in whole numbers for which $0\le u\le v\le n$.
We expand $n+1$ into a product of prime factors and group the
square part of it $n+1=a^2b$, so that $b$ to have not repeated prime
factors.
As $n+1$ divides $u^2$ it follows that  $u=abc$.
We obtain the following inequalities
\begin{equation}
0\le abc\le bc^2+2\le a^2b-1,\quad a,b,c\ge 0,
\end{equation}
which can be rewritten as
\begin{equation}
\left|\begin{array}{rl}bc(a-c)\le 2\\ b(a^2-c^2)\ge3\\ a^2b\ge 2.
\end{array}\right.
\end{equation}
The last implies $b>0$, $a>c$. There must be considered two cases:
\begin{enumerate}
\item $c=0$. This is a solution with $p=q=1$ for arbitrary $n$.
Such an extension can be realized by $r_{n+1}= r_nr_{n-1}r_n$.
\item $c>0$. We have the following system of inequalities
\begin{equation}
0<c<a\le\frac{2}{bc}+c,\quad b>0
\end{equation}
$a,b,c\in\Z_+$ therefore $\frac{2}{bc}\ge 1$. We obtain the following
solutions:
\begin{enumerate}
\item $b=c=1$, $a=2$. It yields $p=\frac{5}{2},q=\frac{1}{2}$ which is not a
solution in whole numbers.
\item $b=c=1$, $a=3$. This is a solution with $p=3,q=0,n=8$. This arrangement
represents redundant generators of $E_8$ (in the minimal realization).
The fifth reflection can be expressed through the others $r_5=gr_9g^{-1}$, where
$g=r_6r_9r_6r_7r_6r_8r_4r_6r_9r_6r_3r_7r_2r_8r_1r_6$.
\item $b=1,c=2,a=3$. This solution gives $p=6,q=0,n=8$. Again the obtained
graph represent redundant generators of $E_8$, which is seen by the identity
$r_6=gr_9g^{-1}$, where $g=r_9r_1r_5r_2r_4r_9r_3r_1r_8r_2r_7$.
\item $b=2,c=1,a=2$. This gives $p=4,q=0,n=7$ and the extension is a degenerate
graph of $E_7$. One of the reflections can be expressed through the remaining ones
$r_5=gr_9g^{-1}$, where $g=r_8r_4r_3r_6r_2r_7r_1r_4$.
\end{enumerate}
\end{enumerate}
The above results imply that all non-degenerate extensions of
$\Gamma_0(A_n)$ without degenerate subgraphs must have $q=0$.
If $p=1$ or $p=n$ the extension generates $A_{n+1}$,
if $p=2$ or $p=n-1$ it generates $D_{n+1}$, and if
$p=3$ or $p=n-2$ $(n<8)$ it generates $E_{n+1}$.

Next we consider the extensions of the universal graph of $D_n$. Using the
same block matrices the following cases must be examined
\begin{equation}\label{Dext1}
\det\begin{pmatrix}2&a_3^t&0&0\\a_3&B_1&a_3&a_1\\0&a_3^t&2&0\\0&a_1^t&0&2
\end{pmatrix}=4(2-p-q)\,,\quad a_3^t=(\overset{n-2}{\overbrace{1,\dotsc,1}})
\end{equation}
\begin{equation}\label{Dext2}
\det\begin{pmatrix}2&a_3^t&0&0\\a_3&B_1&a_3&a_1\\0&a_3^t&2&1\\0&a_1^t&1&2
\end{pmatrix}=8-n-8q,
\end{equation}
\begin{equation}\label{Dext3}
\det\begin{pmatrix}2&a_3^t&0&1\\a_3&B_1&a_3&a_1\\0&a_3^t&2&0\\1&a_1^t&0&2
\end{pmatrix}=8-n-8p,
\end{equation}
\begin{equation}\label{Dext4}
\det\begin{pmatrix}2&a_3^t&0&1\\a_3&B_1&a_3&a_1\\0&a_3^t&2&1\\1&a_1^t&1&2
\end{pmatrix}=4(3+p-n-3q),
\end{equation}
\begin{equation}\label{Dext5}
\det\begin{pmatrix}2&a_3^t&0&-1\\a_3&B_1&a_3&a_1\\0&a_3^t&2&1\\-11&a_1^t&1&2
\end{pmatrix}=4(1-p-q),
\end{equation}
\begin{equation}\label{Dext6}
\det\begin{pmatrix}2&a_3^t&0&-1\\a_3&B_1&a_3&a_1\\0&a_3^t&2&-1\\-11&a_1^t&-1&2
\end{pmatrix}=4(3+q-n-3p).
\end{equation}

The extension \eqref{Dext1} is degenerate if $p=2,q=0$ or $p=q=1$. The first
case is realized with $r_{n+1}= r_{n-1}r_1r_{n-2}r_nr_{n-2}r_1r_{n-1}$
and the second with $r_{n+1}= r_{n-1}r_{n-2}r_{n-1}$. It is non-degenerate
and doesn't contain degenerate subgraphs only if $p=0,q=1$ or $p=1,q=0$ giving
a $D_{n+1}$ arrangement.

The extension \eqref{Dext2} is degenerate only for $n=8,q=0$. In the same way
\eqref{Dext3} is degenerate only for $n=8,p=0$. These two coincide as
reflection arrangements under permutation $(1\,n)$ of the indices of their
reflections. The first
can be realized by $r_7=fgr_8g^{-1}f^{-1}$, where
\begin{equation}
g=r_1hr_9h^{-1}r_2r_8r_3r_1r_4hr_9h^{-1}r_5r_1r_6
\end{equation}
and 
\begin{equation}
\begin{array}{llrl}
f=\Id,& h=r_8&\mbox{if}&p=0\\
f=r_8r_1,& h=r_8&\mbox{if}&p=1\\
f=\Id,& h=r_5r_1r_4r_8r_3r_1r_2r_8&\mbox{if}&p=2\\
f=r_8r_1,& h=r_6r_1r_5&\mbox{if}&p=3\\
f=\Id,& h=r_3r_1r_2r_8&\mbox{if}&p=4\\
f=r_8r_1,& h=r_6r_1r_5r_8r_4r_1r_3&\mbox{if}&p=5\\
f=\Id,& h=\Id&\mbox{if}&p=6\\
\end{array}
\end{equation}
The extension \eqref{Dext2} is admissible and non-degenerate
if $q=0,n<8$ giving $D_{n+1}(n<5)$ or $E_{n+1}(5\le n\le 7)$.

The extension \eqref{Dext4} is degenerate for $q=0,p=n-3$. It can be realized
by $r_{n+1}= r_1r_2r_nr_2r_1$. It is non-degenerate without degenerate
principal minors if $q=0,p=n-2$ giving $\Gamma_0(D_{n+1})$.

The extension \eqref{Dext5} is degenerate for $p=0,q=1$ or
$p=1,q=0$. The first case is realized by $r_{n+1}= r_{n-1}r_nr_{n-1}$
and the second by $r_{n+1}= r_{n-1}r_1r_{n-1}$. It is non-degenerate without
degenerate principal minors if $p=q=0$ giving a $D_{n+1}$ arrangement.

The last extension \eqref{Dext6} is equivalent to~\eqref{Dext4}.

\begin{figure}[htb]
\begin{center}\includegraphics{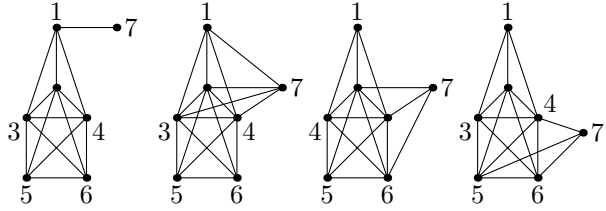}\end{center}
\caption{Degenerate extensions of $\Gamma_0(E_6)$} \label{fig15}
\end{figure}

We come to extensions of graphs, representing generators of the exceptional
groups $E_6,E_7,E_8$. Non-degenerate admissible extensions of
$\Gamma_0(E_6)$ fall in the orbits $E_7^{(k)}$ as we have seen,
while degenerate ones
are realized by $r_7=gr_5g^{-1}$, where $g=r_1r_4r_6r_2r_1r_3$,
$r_4r_6r_2r_1r_3$, $r_6r_2r_1r_3$, $r_2r_1r_3$ respectively for the graphs in
Fig.\ref{fig15}.

\begin{figure}[htb]
\begin{center}\includegraphics{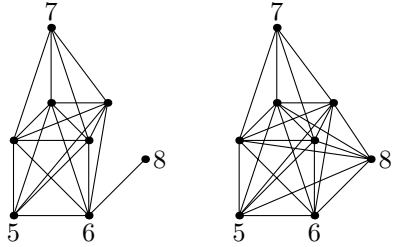}\end{center}
\caption{Degenerate extensions of $\Gamma(D_6\subset E_7,5)$}\label{fig16}
\end{figure}
The orbits of the braid group on non-degenerate configurations of generating
reflections in the group $W(E_7)$ have two ``universal graphs''
$\Gamma(D_6\subset E_7,5)$ and  $\Gamma(D_6\subset E_7,4)$. All their
extensions by one vertex are either degenerate or represent reflections,
generating $W(E_8)$. The admissible degenerate extensions of
$\Gamma(D_6\subset E_7,5)$ can be realized by $r_8=gr_7g^{-1}$,
where $g=r_6r_4r_7r_3r_6r_2r_5r_1$
for the first graph and $g=r_4r_7r_3r_6r_2r_5r_1$ for the second graph
in Fig.\ref{fig16}. The graphs are symmetric with respect to the unindexed
vertices, which must be indexed by the remaining numbers from 1 to 8.
The other ``universal graph'' $\Gamma_0'(E_7)=\Gamma(D_6\subset E_7,4)$
allow only non-degenerate admissible extensions which generate $E_8$.

For the group $W(E_8)$ there are three ``universal''
graphs $\Gamma(D_7\subset E_8,6)$, $\Gamma(D_7\subset E_8,5)$, and the graph
$\Gamma_0(E_8^{(9)})$ from Fig.\ref{fig14}.
\begin{figure}[htb]
\begin{center}\includegraphics{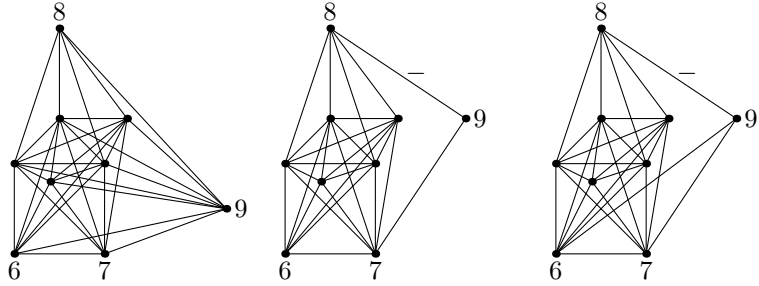}\end{center}
\caption{Degenerate extensions of $\Gamma(D_7\subset E_8,6)$}\label{fig18}
\end{figure}
The admissible extensions of $\Gamma(D_7\subset E_8,6)$ 
are degenerate and can be realized by $r_9=gr_8g^{-1}$,
where
\begin{equation*}
\begin{array}{rcll}
g&=&r_1r_7r_5r_6r_4r_8r_3r_7r_2r_6r_1&\mbox{for the first,}\\
g&=&r_8r_5r_6r_4r_8r_3r_7r_2r_6r_1&\mbox{for the second,}\\
g&=&r_7r_5r_6r_4r_8r_3r_7r_2r_6r_1&\mbox{for the third}
\end{array}
\end{equation*}
graph in Fig.\ref{fig18}.

\begin{figure}[htb]
\begin{center}\includegraphics{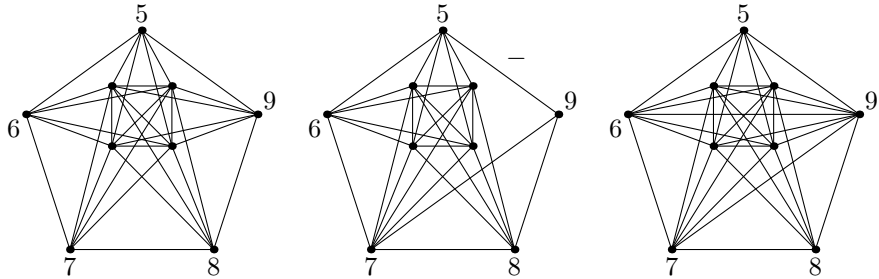}\end{center}
\caption{Degenerate extensions of $\Gamma(D_7\subset E_8,5)$}\label{fighh2}
\end{figure}

All admissible extensions of $\Gamma(D_7\subset E_8,5)$
are degenerate. They can be realized by $r_9=gr_7g^{-1}$, where
\begin{equation*}
\begin{array}{rcll}
g&=&r_8r_7r_4r_8r_3r_5r_2r_6r_1&\mbox{for the first,}\\
g&=&r_7r_4r_8r_3r_5r_2r_6r_1&\mbox{for the second,}\\
g&=&r_7r_5r_4r_8r_3r_5r_2r_6r_1&\mbox{for the third}
\end{array}
\end{equation*}
graphs in  Fig.\ref{fighh2}.

The last ``universal graph'' $\Gamma_0(E_8^{(9)})$
from Fig.\ref{fig14} does not allow admissible extensions.

\begin{cor}
In the minimal realization, all admissible extensions of the universal
arrangements in finite Coxeter groups represent reflections in finite
Coxeter groups.
\end{cor}

\section{General arrangement matrix in a finite orbit}

\subsection{Degenerate arrangements in finite orbits}

An obvious way to obtain degenerate arrangements with finite orbits of
the braid group is to take the generators of a finite group and append
reflections from the same group. As the group is finite these sets
of overdetermined generators are finite. We know from corollary~\ref{realiz}
that there are non-isomorphic realizations of degenerate arrangement matrices,
which gives us a method to construct infinite reflection groups with finite
$B$-orbits. A stronger statement that all groups with the property of having
finite $B$-orbits are obtained in this way is also valid.

\begin{thm}
Any arrangement with positive semi-definite matrix in a finite orbit of the
braid group can be realized as an overdetermined system of generators of
finite Coxeter group.
\end{thm}
\begin{proof}
A positive semi-definite matrix may have only non-negative principal minors.
For an arrangement matrix this means that all subarrangements must have
positive semi-definite matrices. As stated in lemma~\ref{subgraph} the orbit of
an arrangement can be finite only if all of its subarrangements have finite orbits.

Let ${Ar}=\{r_{i_1},r_{i_2},\cdots,r_{i_k}\}$ form a non-degenerate
subsystem of maximal rank with ordered indices $i_1<i_2<\dotsb<i_k$. One may consider
$\widetilde{Ar}=\{r_1,r_2,\cdots,r_n\}$ as an extension of ${Ar}$ by $n-k$
reflections.

The braid $\sigma_j^{-1}$ decreases the index of $r_{j+1}$ by one leaving
$r_l,\,l>j+1$ unchanged, hence
\begin{equation}
\sigma_{k}^{-1}\sigma_{k+1}^{-1}\dotsm\sigma_{i_k-1}^{-1}
\sigma_{k-1}^{-1}\sigma_{k}^{-1}\dotsm\sigma_{i_{k-1}-1}^{-1}
\dotsm\sigma_2^{-1}\sigma_3^{-1}\dotsm\sigma_{i_1-1}^{-1}
\end{equation}
 will transform
$\widetilde{Ar}$ to an arrangement in which the maximal non-degenerate
subsystem is formed from the first $k$ reflections.
Considering all possible extensions by one reflection of the universal
arrangements in every finite Coxeter group we proved that all
arrangements with only non-degenerate sub-arrangements and finite
$B$-orbits generate finite Coxeter groups at least in their minimal
realization in the sense of corollary~\ref{realiz}.
In most cases of degenerate extensions $r_{k+1}$ was expressed through
$r_1,r_2,\cdots,r_k$.
in the case of extensions of $\Gamma_0(D_n)$ to degenerate configurations of
$B_n$ or $\Gamma_0(D_4),\Gamma_0(B_4)$ to $F_4$, or $\Gamma_0(D_8)$ to $E_8$,
or $\Gamma_0(A_8)$ to $E_8$ and $\Gamma_0(A_7)$ to $E_7$
$r_i$ for some $i<k+1$ was expressed through $r_1,r_2,\dotsc,\hat{r}_i,\dotsc,r_{k+1}$.
This difference reflects the following inclusions of Coxeter systems from
the same dimension:
\begin{equation}
D_n\subset B_n,\quad D_4\subset B_4\subset F_4,\quad D_8\subset E_8\supset A_8,
\quad A_7\subset E_7
\end{equation}
These are the only inclusions of irreducible finite Coxeter systems from the
same dimension.

We showed that for all degenerate extensions one of
the reflections belongs to the group generated by the others in the minimal
realization. Now we may take the subarrangement ${Ar}$ with reflections,
generating the whole group and all other $n-k$ reflections will belong to the
same group.

The same argument applies also to reducible arrangements. As for irreducible
extensions of reducible configurations we may always take another irreducible
subsystem and consider its extension. The following inclusions of reducible
Coxeter systems in finite irreducible systems of the same dimension appear:
\begin{gather}
A_1\times A_1\subset B_2,\quad
D_k\times B_{n-k}\subset B_{n},\\
A_1\times A_1\times A_1\subset B_3,\quad
B_k\times B_{n-k}\subset B_{n},\\
B_n\times A_1\subset B_{n+1},\quad
A_1^{\times 4}\subset D_4,\quad
D_k\times D_{n-k}\subset D_{n},\\
A_1\times A_5\subset E_6,\quad A_2\times A_2\times A_2\times\subset E_6\\
A_1\times A_3\times A_3\subset E_7,\quad
A_2\times A_5\subset E_7,\quad A_1\times D_5\subset E_7\\
A_1\times A_2\times A_5\subset E_8,\quad A_1\times A_7\subset E_8,\\
A_4^{\times 2}\subset E_8,\quad A_3\times D_5\subset E_8,\quad
A_1\times E_7\subset E_8\\
A_1\times A_1\times A_1\subset H_3,\quad A_1\times H_3\subset H_4\\
I_2(5)\times I_2(5)\subset H_4,\quad
A_2\times A_2\subset H_4
\end{gather}
The easiest way to obtain these inclusions is to take minimally connected
graphs of degenerate configurations and remove one vertex in all possible ways.
For the Weyl groups these graphs are the extended Dynkin diagrams of
affine Coxeter groups. For the non-crystallographic
systems $H_3,H_4$ can be used the graphs in Fig.\ref{figh23},
which represent degenerate configurations.
\begin{figure}[htb]
\begin{center}\includegraphics{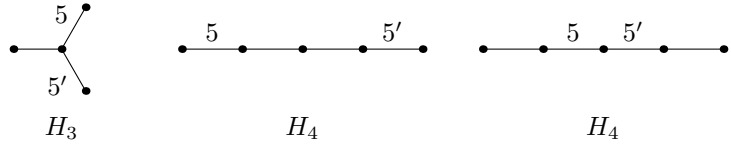}\end{center}
\caption{Minimally connected degenerate arrangements for $H_3,H_4$}
\label{figh23}
\end{figure}
\end{proof}

Although the theorem describes how to be obtained finite
orbits of the braid group on degenerate systems of reflections the actual
determination of these orbits is far from complete. These
orbits may hide additional invariant foliated symplectic
structure as in the case of rank 2 matrices
discussed in~\cite{Al}.

\subsection{The main theorem}

In order to determine all the symmetrized Stokes matrices with finite
orbits under the action of the braid group one needs to consider apart from
positive semidefinite also the indefinite arrangement matrices.
Out attempt to build inductively matrices with finite orbits by adding one
row and column to matrices with proved finite orbit may fail because there are
some invertible $n\times n$ matrices whose all principal minors of size $n-1\times n-1$
are degenerate. Examples are
\begin{equation}
\begin{pmatrix}2&-2&-2\\-2&2&-2\\-2&-2&2\end{pmatrix}
\end{equation}
\begin{equation}
\begin{pmatrix}
2&2\cos\frac{p\pi}{d}&2\cos\frac{q\pi}{d}&2\cos\frac{(p-q)\pi}{d}\\
2\cos\frac{p\pi}{d}&2&2\cos\frac{(p-q)\pi}{d}&2\cos\frac{q\pi}{d}\\
2\cos\frac{q\pi}{d}&2\cos\frac{(p-q)\pi}{d}&2&2\cos\frac{p\pi}{d}\\
2\cos\frac{(p-q)\pi}{d}&2\cos\frac{q\pi}{d}&2\cos\frac{p\pi}{d}&2
\end{pmatrix}
\end{equation}
These matrices are indefinite and do not have finite orbits but all their
principal minors, which are the arrangement matrices of their subarrangements
are degenerate and have finite orbits.
The following lemma is essential.

\begin{lem}
An invertible arrangement matrix $B$ with $n$ rows for which all principal
minors of degree $n-1$ are degenerate can always be transformed by a suitable
braid to a matrix without this property.
\end{lem}
\begin{proof}
Recall that a principal minor is a submatrix obtained by deleting rows and
columns with the same numbers. We have
\begin{equation}
A_{ij}={B^{-1}}_{ij}=\frac{\det(B_{pq})_{p\ne i,q\ne j}}{\det(B)},
\end{equation}
so the above property implies $A_{ii}=0$ for every $i$.

The canonical generators of the braid group transform the matrix $B$ in the
following way
\begin{gather}
\sigma_i(B)=K_i(B)\cdot B\cdot K_i(B)\,,\\ K_i(B)=
\begin{pmatrix}\Id_{i-1,i-1}&0&0&0\\0&-B_{i,i+1}&1&0\\0&1&0&0\\0&0&0&
\Id_{n-i-1,n-i-1}&
\end{pmatrix}
\end{gather}
The inverse matrix $A$ is transformed correspondingly
\begin{equation}
\sigma_i(A)=K_i(B)^{-1}\cdot A\cdot K_i(B)^{-1}.
\end{equation}

Assume that for any braid transformation the matrix $B$ preserves its
property of having only degenerate principal minors. The generators of the
braid group transform the entries $B_{i+1,i+k},A_{i+1,i+k}$ for $k\ge 1$ as
\begin{equation}
\begin{array}{rl}
\sigma_i(B)_{i+1,i+k}&=B_{i,i+k}\\
\sigma_i(A)_{i+1,i+k}&=A_{i,i+k}+B_{i,i+1}A_{i+1,i+k}
\end{array}
\end{equation}
\begin{equation}
\begin{array}{rl}
\sigma_i^{-1}(B)_{i+1,i+k}&=B_{i,i+k}-B_{i,i+1}B_{i+1,i+k}\\
\sigma_i^{-1}(A)_{i+1,i+k}&=A_{i,i+k}
\end{array}
\end{equation}
The diagonal entries of $A$ are zero and must remain zero after the action of
any braid. We will prove by induction that this implies $A_{ij}B_{ij}=0$. We
have $A_{ii}B_{ii}=0$. Assume it is true that $A_{i,i+k-1}B_{i,i+k-1}=0$ must
hold for every $i$ in order $A_{ii}$ to remain zero under any braid.
Acting with $\sigma_i$ we obtain
\begin{equation}
\left|\begin{array}{rl}
A_{i+1,i+k}B_{i+1,i+k}&=0\\
B_{i,i+k}A_{i,i+k}+B_{i,i+1}B_{i+1,i+k}&=0\\
B_{i,i+k}A_{i,i+k}-B_{i,i+1}A_{i+1,i+k}&=0\end{array}\right.
\end{equation}
which implies $B_{i,i+k}A_{i,i+k}=0$. By induction on $k$ we find that for
diagonal entries of $A$ to remain zero under any braid it is necessary
to have $A_{ij}B_{ij}=0$ which is an absurd as
\begin{multline}
\det(B)=\sum_{j=1}^n(-1)^{i+j}B_{ij}\det(B_{pq})_{p\ne i,q\ne j}\\
=\sum_{j=1}^n(-1)^{i+j}B_{ij}\det(B)A_{ij}=0\,.
\end{multline}
We conclude that either the matrix $B$ is degenerate or there is a braid
transforming it such that there is a non-degenerate principal minor
of degree $n-1$.
\end{proof}

From an arbitrary $n\times n$ arrangement matrix $B$ can be constructed a chain
of its principal minors $B=B_n\supset B_{n-1}\supset\dotsb\supset B_1=(2)$
in the following way. Choose a principal minor of degree $n-1$ with maximal
rank and call it $B_{n-1}$ and its rank $r_{n-1}$. Clearly $r_n\ge r_{n-1}$.
Continue by choosing a principal minor in $B_{n-1}$ of maximal rank and so on.
We denote by $s_i$ the size of the biggest non-degenerate principal minor in
$B_i$. Here we count also the matrix $B_i$ as a principal minor of itself.

\begin{thm}
If in the chain constructed above exists a number $i$ such that $s_i>s_{i-1}
+1$, there is a braid transforming $B$ to $B'$ for which $s'_i=s'_{i-1}+1$.
\end{thm}
\begin{proof}
We consider the matrix $B_{s_i}$ which is non-degenerate by definition. It is
contained in $B_j$, $j\ge s_i$ therefore $i-1<s_i$.On the other hand $s_i\le i$
which implies $s_i=i$. We have $s_{i-1}<s_i-1=i-1$ therefore while the matrix
$B_i$ is non-degenerate all its principal minors of size $i-1$ must be
degenerate. The previous lemma concludes the proof.
\end{proof}

We obtain that $s_i$ takes only values $s_{i-1}$ or $s_{i-1}+1$ for some matrix
in the same orbit of the braid group.

\begin{thm}
All arrangement matrices with finite orbits are either non-de\-gen\-er\-ate
corresponding to reflections generating finite groups or their extensions
with reflections from the same group.
\end{thm}
\begin{proof}
We proceed by induction. It is proved for the case of 3 by 3 matrices.
Assume true for $n\times n$ matrices. Any $n+1\times n+1$ matrix $B$ which is
non-degenerate contains an $n\times n$ non-degenerate subarrangement or can
be made so by a suitable braid. There were considered all extensions of
non-degenerate arrangements generating finite Coxeter groups and it was shown
that they must generate again finite reflection group in order to have finite
orbit. Now let the matrix $B$ be degenerate with maximal non-degenerate
principal minor $B_s$. By induction hypothesis $B_s$ is an arrangement
generating finite Coxeter group. All degenerate extensions by one reflection
of $B_s$ must have the new reflection in the group generated by the other or
otherwise the orbit of the extended matrix will be infinite. It follows
that all reflections in $B$ must belong to the group generated by $B_s$.
\end{proof}

\section{Conclusion}
The classification of the orbits is unfinished. It will be interesting to find
if it is possible to linearize in a uniform way the action of the braid group
as it was done for the rank two arrangements. One may expect that there
will be some hidden structures in analogy with the symplectic structure, which
was found in the studied rank two case.

It is appealing how far can be extended the interpretation of configurations
with higher degeneracy. Whether these can be used for classification of the
quasi-periodic tilings? How must the definition of abstract presentation of
Coxeter groups be extended to include groups generated by reflections with
such arrangement matrices?

The action of the braid group on pseudoreflections generating finite unitary
groups is considered in~\cite{Ph,Ph2}. The combinatorics of these complex
reflections is not well understood. One way to tackle the problem of
absence of notion about simple roots is to consider all possible $n$-tuples of
pseudoreflections generating finite groups, where the results of the present
work would be helpful.
\vskip3ex
\noindent{\large\bf Acknowledgements}
This article is based on the PhD thesis of the author.
I am grateful to my supervisor Boris Dubrovin for introducing me
to this field and for his guidance and encouragement. I wish to thank to
all members of the Mathematical Physics sector at SISSA
for their hospitality.

\newpage
\addcontentsline{toc}{section}{Appendices}
\addcontentsline{toc}{subsection}{Table A}
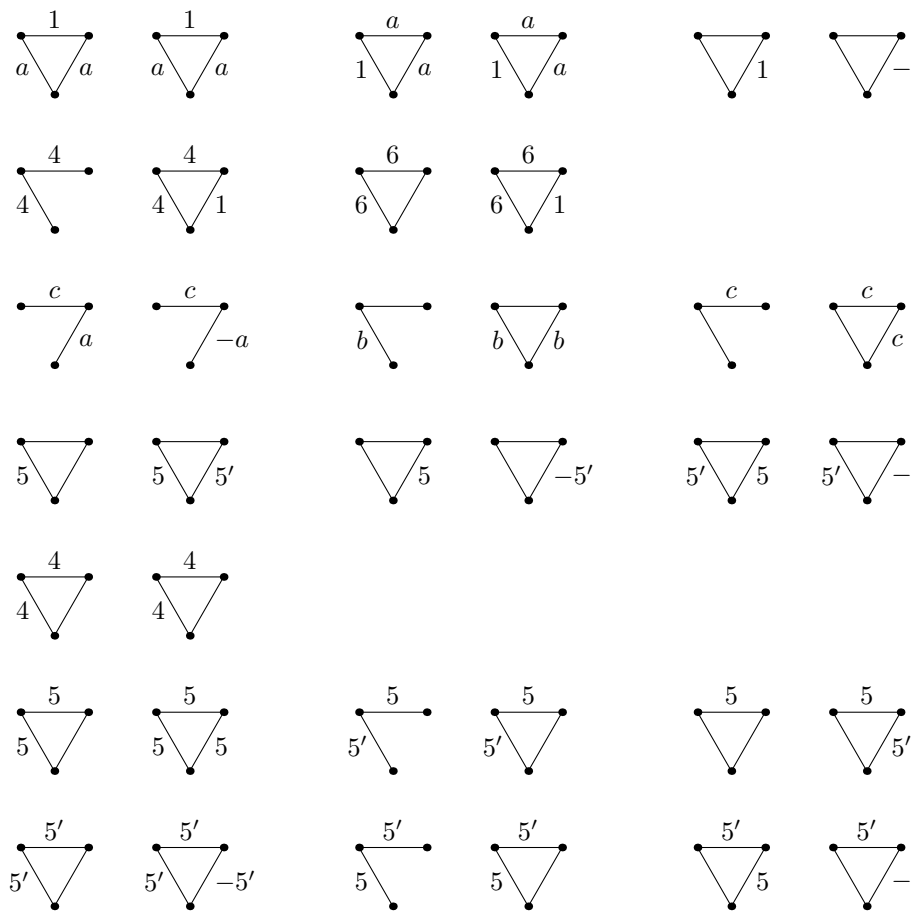
\centerline{\hfil Table A\hfil}
\centerline{\bf Transformations of graphs under the change
$r_j\mapsto r_ir_jr_i$}\vskip3ex
\centerline{\includegraphics{table.1}}
\vskip3ex

Only the edges ending to the $j$-th vertex are affected. The transformation
has period 2. In the table are given pairs of interchanging graphs. The
$i$-th vertex is the upper-left corner of the triangle and the $j$-th is the
upper-right. The changes from this table must be applied to all pairs
$\{j,k\}$ for which the $k$-th vertex is joined to the $i$-th or $j$-th.

\newpage
\addcontentsline{toc}{subsection}{Table B}
\centerline{\hfil Table B\hfil}\vskip1cm
\begin{tabular}{cccc}
Family & Orbit & Graphs & $\det(C-\Id x)$\\ \hline
A & 1 &\includegraphics{figh.1} & 
$\left(x^2-2x\cos(\frac{\pi}{15})+1\right)\left(x^2-2x\cos(\frac{11\pi}{15})+1\right)$\\[2ex]
  & 2 &\lower.5cm\hbox{\includegraphics{figh.2}}
$q_<=2$ & 
$\left(x^2-2x\cos(\frac{\pi}{5})+1\right)^2$\\[4ex]
B & 1 &\includegraphics{figh.3} & 
$\left(x^2-2x\cos(\frac{7\pi}{15})+1\right)\left(x^2-2x\cos(\frac{13\pi}{15})+1
\right)$\\[2ex]
  & 2 &\lower.5cm\hbox{\includegraphics{figh.4}} $q_<=2$ &
$\left(x^2-2x\cos(\frac{3\pi}{5})+1\right)^2$\\[4ex]
C & 1 &\includegraphics{figh.5} & 
$\left(x^2-2x\cos(\frac{3\pi}{10})+1\right)\left(x^2-2x\cos(\frac{7\pi}{10})+1
\right)$\\[2ex]
  & 2 &\includegraphics{figh.6} & 
$\left(x^2-2x\cos(\frac{4\pi}{15})+1\right)\left(x^2-2x\cos(\frac{14\pi}{15})+1
\right)$\\[2ex]
D & 1 &\includegraphics{figh.7} & 
$\left(x^2-2x\cos(\frac{\pi}{10})+1\right)\left(x^2-2x\cos(\frac{9\pi}{10})+1
\right)$\\[2ex]
  & 2 &\includegraphics{figh.8} & 
$\left(x^2-2x\cos(\frac{2\pi}{15})+1\right)\left(x^2-2x\cos(\frac{8\pi}{15})+1
\right)$\\[2ex]
E & 1 &\includegraphics{figh.9} & $\Phi_{12}(x)$\\[2ex]
  & 2 &\lower.5cm\hbox{\includegraphics{figh.10}} $q_<=2$& 
$\Phi_{10}(x)$ 
\\[4ex]
  & 3 &\lower.5cm\hbox{\includegraphics{figh.11}} $q_<=2$& $
\Phi_6(x)^2$\\
\end{tabular}
\[5'=\frac{5}{2}\]

\newpage
\addcontentsline{toc}{subsection}{Table C}
\centerline{\hfil Table C\hfil}\vskip.5cm
Minimal realizations of degenerate extensions of $H_3$ configurations.
\vskip.5cm
\includegraphics{figh.25}
\newpage
\addcontentsline{toc}{subsection}{Table D}
\centerline{\hfil Table D\hfil}\vskip.5cm
Admissible non-degenerate extensions of universal graphs of $H_4$
\vskip.5cm
\hbox to\textwidth{\includegraphics[width=\textwidth]{fighh.1}}
\newpage
\addcontentsline{toc}{subsection}{Table E}
\centerline{\hfil Table E\hfil}\vskip.5cm
Minimal realizations of degenerate extensions of $H_4$ configurations.
\vskip.5cm
\hbox to\textwidth{\includegraphics{figur.26}}
\hbox to\textwidth{\hfil continues on next page\dots}
\newpage
\centerline{\hfil Table E\hfil}\vskip.5cm
\vskip.5cm
\hbox to\textwidth{\includegraphics{figur.27}}
\hbox to\textwidth{\hfil continues on next page\dots}
\newpage
\centerline{\hfil Table E\hfil}\vskip.5cm
\vskip.5cm
\hbox to\textwidth{\includegraphics{figur.271}}
\hbox to\textwidth{\hfil continues on next page\dots}
\newpage
\centerline{\hfil Table E\hfil}\vskip.5cm
\vskip.5cm
\hbox to\textwidth{\includegraphics{figur.272}}
\hbox to\textwidth{\hfil continues on next page \dots}
\newpage
\centerline{\hfil Table E\hfil}\vskip.5cm
\vskip.5cm
\hbox to\textwidth{\includegraphics{figur.29}}
\hbox to\textwidth{\hfil continues on next page\dots}
\newpage
\centerline{\hfil Table E\hfil}\vskip.5cm
\vskip.5cm
\hbox to\textwidth{\includegraphics{figur.30}}
\hbox to\textwidth{\hfil continues on next page \dots}
\newpage
\centerline{\hfil Table E\hfil}\vskip.5cm
\vskip.5cm
\hbox to\textwidth{\includegraphics{figur.31}}

\bibliography{biblio.bib}
\bibliographystyle{plain}

\end{document}